\documentclass[acmsmall, authorversion]{acmart}

\usepackage{amsfonts}
\usepackage{bm}
\usepackage{dirtytalk}
\usepackage{multirow}
\usepackage{subcaption}
\usepackage{colortbl}
\usepackage{xspace}
\usepackage{hyperref}

\AtBeginDocument{%
  }

\setcopyright{cc}
\setcctype{by}
\acmJournal{PACMCGIT}
\acmYear{2025} \acmVolume{8} \acmNumber{1} 
\acmMonth{5} \acmPrice{}\acmDOI{10.1145/3728311}

\keywords{Virtual Reality, 3D Gaussian Splatting, Point-based Rendering, Radiance Fields, Real-time Graphics, Real-time Rendering}

\acmISBN{978-1-4503-XXXX-X/18/06}

\acmSubmissionID{23}

\usepackage[ruled]{algorithm2e} 

\SetAlFnt{\small}
\SetAlCapFnt{\small}
\SetAlCapNameFnt{\small}
\SetAlCapHSkip{0pt}

\definecolor{xuechang_color}{rgb}{.6,.4,.05}
\definecolor{michael_color}{rgb}{0,0.35,0}
\definecolor{lukas_color}{rgb}{.66,.25,0.85}
\definecolor{markus_color}{rgb}{0,0.35,0.35}
\definecolor{bernhard_color}{rgb}{0.35,0.35,0}

\definecolor{tab_color}{rgb}{0.0,0.5,0.5}

\definecolor{primary_color}{HTML}{007777}
\definecolor{secondary_color}{HTML}{CC5522}

\newcommand{\ifcommentsenabled}[1]{}

\definecolor{edited_color}{rgb}{.0,.1,.7}
\newcommand{\new}[1]{{#1}} 
\newcommand{\newnew}[1]{{#1}}

\definecolor{revised_color}{rgb}{.1,.7,.1}
\newcommand{\revised}[2]{{#1}} 
\newcommand{\newrevised}[2]{{#1}}

\renewcommand{\vec}[1]{\mathbf{#1}} 

\newcommand{\FLIP}{\protect\reflectbox{F}LIP\xspace}

\newcommand{\gs}{3DGS\xspace}
\newcommand{\sigmatwod}{\Sigma_{\text{2D}}}
\newcommand{\minigs}{Mini-Splatting\xspace}
\newcommand{\ours}{Ours\xspace}
\newcommand{\stp}{StopThePop\xspace}

\newcommand{\ie}{\textit{i.e.}\xspace}
\newcommand{\eg}{\textit{e.g.}\xspace}
\newcommand{\cf}{cf.\xspace}

\newcommand{\mue}{\bm{\mu}}
\newcommand{\muetd}{\bm{\mu}_{\text{2D}}}

\begin{document}
\title{VRSplat: Fast and Robust Gaussian Splatting for Virtual Reality}


\author{Xuechang Tu}
\orcid{0000-0002-1338-1653}
\affiliation{%
     \institution{Peking University}
     \country{China}
}
\email{cmxrynp@gmail.com}

\author{Lukas Radl}
\orcid{0009-0008-4075-5877}
\email{lukas.radl@tugraz.at}
\affiliation{%
    \institution{Graz University of Technology}
    \country{Austria}
}

\author{Michael Steiner}
\orcid{0009-0008-7430-6922}
\email{michael.steiner@tugraz.at}
\affiliation{%
    \institution{Graz University of Technology}
    \country{Austria}
}

\author{Markus Steinberger}
\orcid{0000-0001-5977-8536}
\affiliation{%
    \institution{Graz University of Technology}
    \country{Austria}
}
\affiliation{%
    \institution{Huawei Technologies}
    \country{Austria}
}
\email{steinberger@tugraz.at}

\author{Bernhard Kerbl}
\orcid{0000-0002-5168-8648}
\email{kerbl@cg.tuwien.ac.at}
\affiliation{%
    \institution{Carnegie Mellon University}
    \country{USA}
}
\author{Fernando de la Torre}
\orcid{0000-0002-7086-8572}
\email{ftorre@andrew.cmu.edu}
\affiliation{%
    \institution{Carnegie Mellon University}
    \country{USA}
}

\renewcommand\shortauthors{Tu, X. et al}
\begin{teaserfigure}
\begin{center}
  \includegraphics[width=\textwidth]{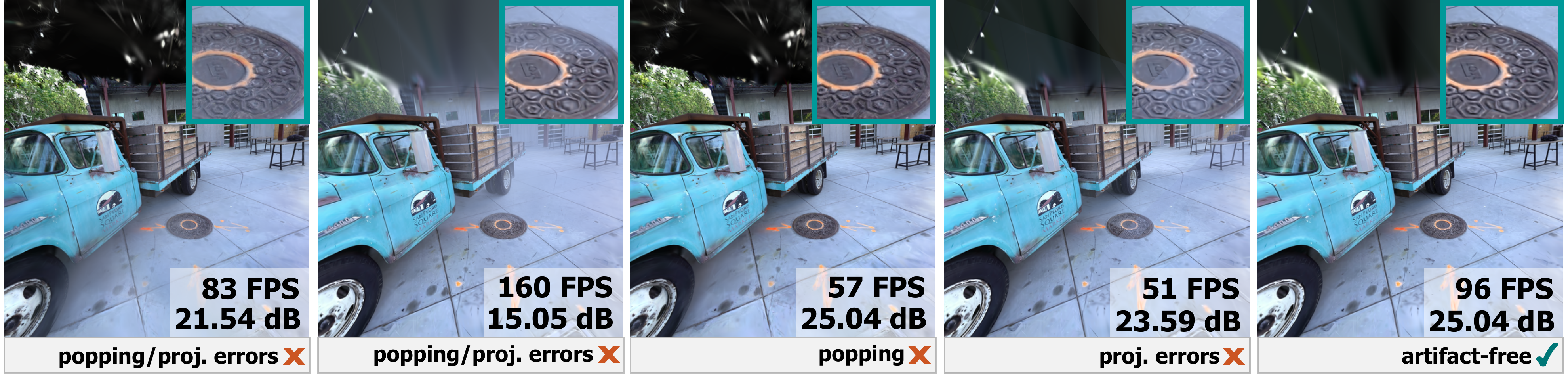}
\begin{minipage}[t]{0.195\textwidth}
    \centering
    \footnotesize
    \text{3DGS}
\end{minipage}
\begin{minipage}[t]{0.195\textwidth}
    \centering
    \footnotesize
    \text{Mini-Splatting}
\end{minipage}
\begin{minipage}[t]{0.195\textwidth}
    \centering
    \footnotesize
    \text{Optimal Projection}
\end{minipage}
\begin{minipage}[t]{0.195\textwidth}
    \centering
    \footnotesize
    \text{StopThePop}
\end{minipage}
\begin{minipage}[t]{0.195\textwidth}
    \centering
    \footnotesize
    \text{Ours}
\end{minipage}
\end{center}
  \caption{
\revised{For a large field-of-view as commonly used in VR}{For a large field-of-view (as used in VR)}, 3DGS's projection error produces elongated or large Gaussians that can cause cloudy appearance, leading to significantly worse PSNR \newrevised{values}{metrics}.
Optimal Projection faithfully projects these Gaussians, but incurs a performance cost and still suffers from immersion-breaking popping artifacts.
Mini-Splatting can drastically reduce model size and rendering times, but resolving remaining popping artifacts (StopThePop) introduces significant overhead.
Our full method elegantly combines these techniques with an efficient foveated rendering routine and provides an immersive and artifact-free VR experience with framerates above 72FPS rendered on an NVIDIA RTX 4090 and displayed on a Meta Quest 3.
}
\label{fig:teaser}
\end{teaserfigure}
\begin{abstract}

3D Gaussian Splatting (3DGS) has rapidly become a leading technique for novel-view synthesis, providing exceptional performance through efficient software-based GPU rasterization. Its versatility enables real-time applications, including on mobile and lower-powered devices. 
However, 3DGS faces key challenges in virtual reality (VR): 
(1) temporal artifacts, such as popping during head movements, 
(2) projection-based distortions that result in disturbing and view-inconsistent floaters, and 
(3) reduced framerates when rendering large numbers of Gaussians, falling below the critical threshold for VR. 
Compared to desktop environments, these issues are drastically amplified by large field-of-view\new{, constant head movements,} and high resolution of head-mounted displays (HMDs).
In this work, we introduce VRSplat: we combine and extend several recent advancements in 3DGS to address challenges of VR holistically. 
We show how the ideas of Mini-Splatting, StopThePop, and Optimal Projection can complement each other, by modifying the individual techniques and core 3DGS rasterizer. 
Additionally, we propose an efficient foveated rasterizer that handles focus and peripheral areas in a single GPU launch, avoiding redundant computations and improving GPU utilization. 
Our method also incorporates a fine-tuning step that optimizes Gaussian parameters based on StopThePop depth evaluations and Optimal Projection.
We validate our method through a controlled user study with 25 participants, showing a strong preference for VRSplat over other configurations of Mini-Splatting. VRSplat is the first, systematically evaluated 3DGS approach capable of supporting modern VR applications, achieving 72+ FPS while eliminating popping and stereo-disrupting floaters. 

\end{abstract}  

\ccsdesc[500]{Computing methodologies~Virtual reality}
\ccsdesc[500]{Computing methodologies~Rasterization}
\ccsdesc[300]{Computing methodologies~Machine learning}

\maketitle


\section{Introduction}

The field of novel-view synthesis has experienced remarkable advancements in recent years, driven by the introduction of Neural Radiance Fields (NeRFs)~\cite{Mildenhall2020NeRF} and the wide array of subsequent developments. 
More recently, 3D Gaussian Splatting (3DGS)~\cite{kerbl3Dgaussians} has introduced a paradigm shift by representing scenes with explicit 3D Gaussian point clouds, rather than implicit volumetric fields. 
3DGS offers not only high visual quality and low training times, but also enables efficient rendering through GPU-based software rasterization or the traditional graphics pipeline.

Capturing and synthesizing novel views of real-world scenes is crucial in a variety of applications, such as e-commerce, visual effects in film production, and immersive experiences. 
Virtual reality (VR) has emerged as a particular focus, with its stringent demands for high framerates and interactivity. 
The impressive performance of 3DGS rasterization makes it well-suited for rendering on low-powered head-mounted displays (HMDs), but several limitations in the original 3DGS implementation can disrupt the immersive experience.

Firstly, the reliance on global primitive sorting during rasterization introduces popping artifacts when the viewpoint changes~\cite{radl2024stopthepop}, which can be particularly jarring in VR due to the constant micro-movements of the head in HMDs.
Secondly, the projection of 3D Gaussians into 2D splats is based on a local affine approximation~\cite{zwicker_ewa_2001}, leading to projection errors that worsen with increasing distance from the image plane center~\cite{huang2024optimal}.
These errors are further amplified in VR due to the wide field-of-view (FOV), causing Gaussians to become distorted or cloud-like, which detracts from the visual experience.
Lastly, the heuristic-driven densification in 3DGS reconstructions often results in excessively large point clouds, making it challenging to maintain interactive framerates since the runtime is directly tied to the number of primitives.

To address these limitations, we introduce VRSplat, a robust and efficient solution for rendering Gaussian splats in VR. 
Our method combines recent advancements in 3DGS in a meaningful way and introduces novel, targeted high-performance optimizations to ensure view-consistent, artifact-free rendering while maintaining the required framerates for VR. 
While VRSplat \revised{can be used with \emph{any} scene reconstruction method that generates a compact set of 3D Gaussians}{remain baseline-agnostic}, we choose \emph{Mini-Splatting}~\cite{fang2024minisplatting} \newrevised{}{as a} due to its reduced primitive count.
We mitigate popping artifacts by employing hierarchical rasterization as seen in StopThePop~\cite{radl2024stopthepop}, and apply Optimal Projection~\cite{huang2024optimal} to eliminate projection artifacts.
However, both methods come with a significant performance cost. 
To counter this, we propose a single-pass foveated rendering technique, which boosts performance and ensures that the recommended frame rate of $\geq 72$ FPS at native HMD resolutions is consistently achieved.
In addition to a novel system design for real-time artifact-free rendering, VRSplat makes the following technical contributions:
\begin{itemize}
    \item A fine-tuning approach that takes a 3DGS scene trained with any (compact) optimization strategy as input---for example Mini-Splatting which reduces the number of Gaussians by up to $10\times$---and optimizes it for both StopThePop rendering and Optimal Projection.
    This approach not only removes large floating artifacts and popping but also significantly improves overall reconstruction quality.
    \item The integration of StopThePop rendering and Optimal Projection within the core 3DGS CUDA renderer.
    This combination enables correct and efficient tile-based culling by accounting for tight bounds from Optimal Projection and providing efficient hierarchical sorting and culling via StopThePop.
    \item An optimized Gaussian Splatting rasterizer that supports rendering tiles at varying resolutions, enabling effective foveated rendering in a single pass. 
    This approach eliminates redundant computations, minimizes kernel launch overhead, reduces memory access during preprocessing and sorting, and improves GPU utilization by handling the entire workload in one go.
\end{itemize}
Finally, this paper contributes a formal validation of the enhanced VR experience provided by VRSplat, via an extensive user study; its evaluation shows overwhelming preference for VRSplat compared to baseline, highlighting our method’s ability to deliver an immersive VR experiences and meet the performance requirements of \revised{smooth VR applications}{modern HMDs}.
\new{We additionally demonstrate significantly improved image quality metrics when adapting the evaluation setup to the requirements of VR rendering (\ie high-resolution, large FOV).}
Fig.~\ref{fig:teaser} illustrates the limitations of the individual techniques we employ, and how, when combined with our single-pass foveated rendering solution, they form a view-consistent VR renderer capable of rendering large 3D scenes.


\section{Related Work}
This section revisits recent work on radiance fields, their applications in VR, and approaches toward interactive rendering on HMDs.

\paragraph{Radiance Fields.}
Novel-view synthesis is a long-standing problem in visual computing.
Neural Radiance Fields (NeRFs)~\cite{Mildenhall2020NeRF} have demonstrated impressive results by combining classical volumetric rendering along view rays combined with large MLPs.
While NeRFs exhibit astonishing visual quality, they suffer from long training and rendering times due to expensive raymarching for each pixel.
Recent advances have focused on accelerating rendering and training~\cite{mueller2022instant, fridovich2022plenoxels, steiner2024nerfcaching}, extended NeRFs to unbounded scenes~\cite{barron2022mipnerf360, barron2023zip} or applied them for various downstream applications, \emph{e.g.} 3D Stylization~\cite{nguyen2022snerf, zhang2023refnpr, radl2024laenerf}.

3D Gaussian Splatting (3DGS)~\cite{kerbl3Dgaussians} thus introduces 3D Gaussians as an alternative scene representation, leveraging efficient software rasterization.
Similarly to NeRF, this resulted in another wave of follow-up work, exploring artifact-free rendering~\cite{Yu2023MipSplatting,radl2024stopthepop,huang2024optimal}, reducing the number of primitives~\cite{fang2024minisplatting, mallick2024taming,fan2023lightgaussian, kheradmand2024mcmc} and rendering of very large datasets~\cite{kerbl2024hierarchical, lin2024vastgaussian}, amongst many other works. 

\paragraph{Radiance Fields for Virtual Reality.}

VR is a natural use case for radiance fields, as it provides more immersive experiences when exploring scenes. 
FoV-NeRF~\cite{deng2022fovnerf} utilizes a gaze-contingent neural representation and foveated rendering to achieve near real-time rendering of NeRF in VR. 
VR-NeRF~\cite{xu_vr-nerf_2023} instead distributes computation across multiple GPUs and leverages an occupancy grid to improve performance. 
However, their trade-off between quality and speed is far from satisfactory.

VR-GS~\cite{jiang2024vrgs} builds an interactive VR setting on top of the original 3DGS renderer that offers intuitive interaction with objects using simulation. 
Our approach could be used as a drop-in replacement to offer higher-quality rendering.
Concurrent work to ours RTGS~\cite{lin2024rtgs} also explores rendering of Gaussians on mobile devices and proposes a foveated rendering solution to achieve immersive framerates.
Our own method, VRSplat, substantiates the ideas presented in a recent poster abstract~\cite{anon2024fastandrobust}.
The contributions of RTGS are complementary to ours: 
they propose a training strategy that leads to scene reconstructions that should increase rendering speeds by reducing the number of overlapping Gaussians.
Their second contribution is a trained, level-of-detail-based foveated rendering strategy that selects a lower number of Gaussians for lower-resolution rendering.
In contrast to RTGS, our fine-tuning procedure is significantly more lightweight due to a single model being used, rendered at different resolution.
Our approach could be combined with RTGS by fine-tuning their output to our model to reduce artifacts and render foveated images in a single pass to further increase VR framerates.

\paragraph{Accelerating Virtual Reality Rendering.}
The characteristics of the human visual system have long been exploited to accelerate rendering.
For instance, due to reduced visual acuity in the periphery, a plethora of works~\cite{guenter2012foveated, patney2016foveated,sun2017perceptually,kaplanyan2019deepfovea,konrad2020gaze,chakravarthula2021gaze,rolff2023interactive,krajancich2023towards,rolff2023vrs} reduce rendering quality in these regions, thereby increasing rendering speed.
Recent work has also leveraged procedural noise generation in foveated rendering, effectively simulating the presence of high-frequency content outside the fovea in a post-processing step~\cite{tariy2022noisebased}.
Our single-pass foveated rendering solution, although related to the aforementioned works, specifically \revised{operates}{operations} on 3DGS's tile-based rasterization to extract the maximum performance.

\begin{figure}
    \centering
    \includegraphics[width=0.4\linewidth]{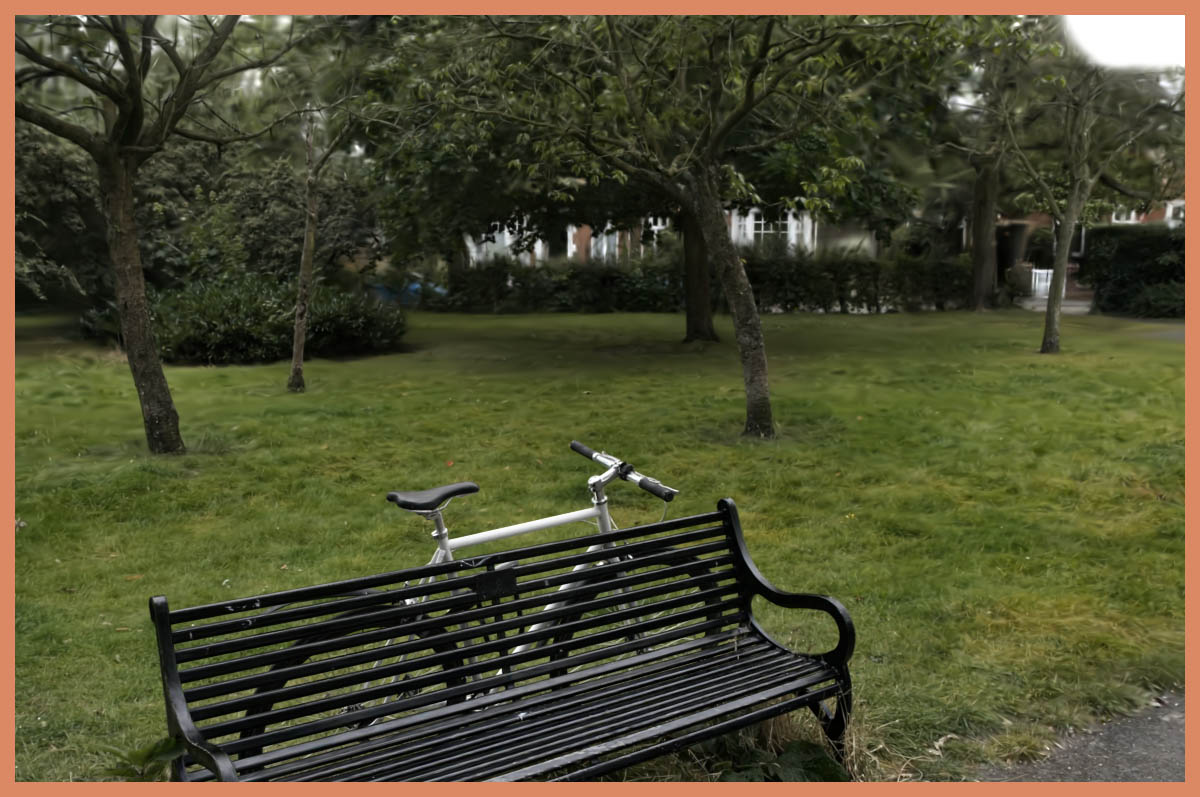}
    \includegraphics[width=0.4\linewidth]{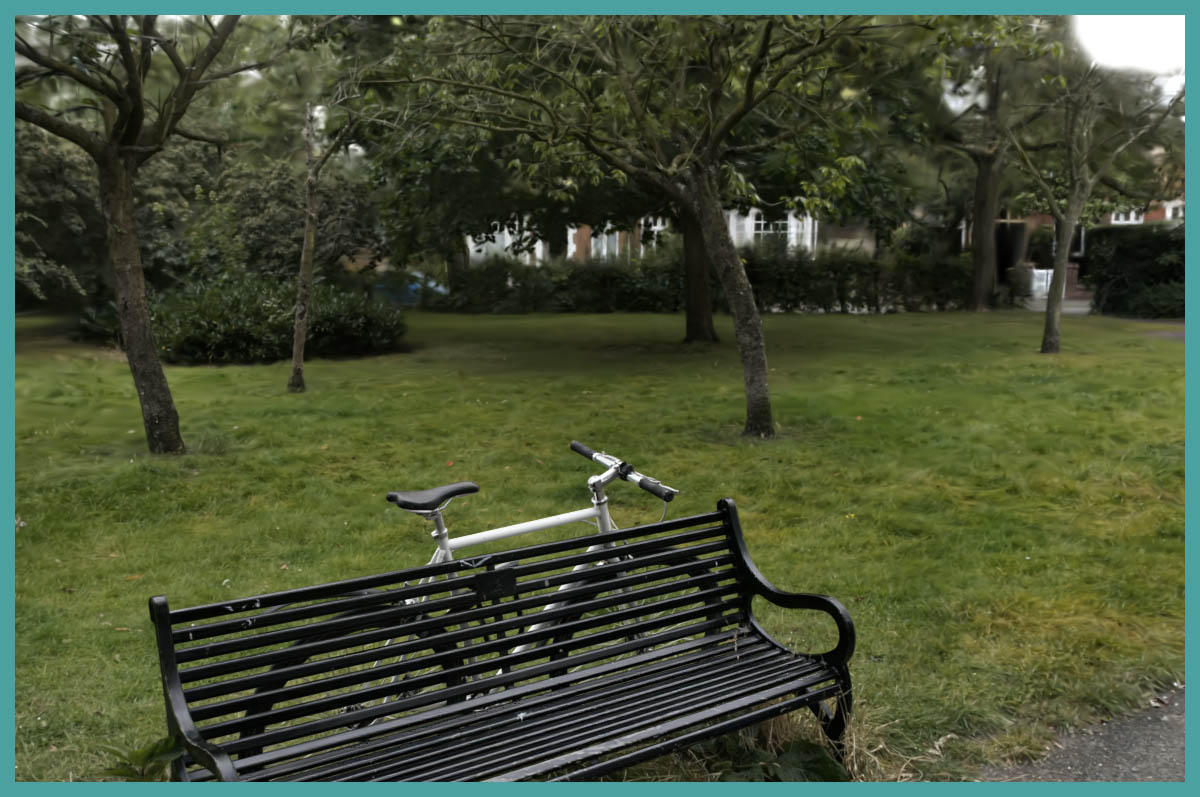}
    \includegraphics[width=0.24\linewidth]{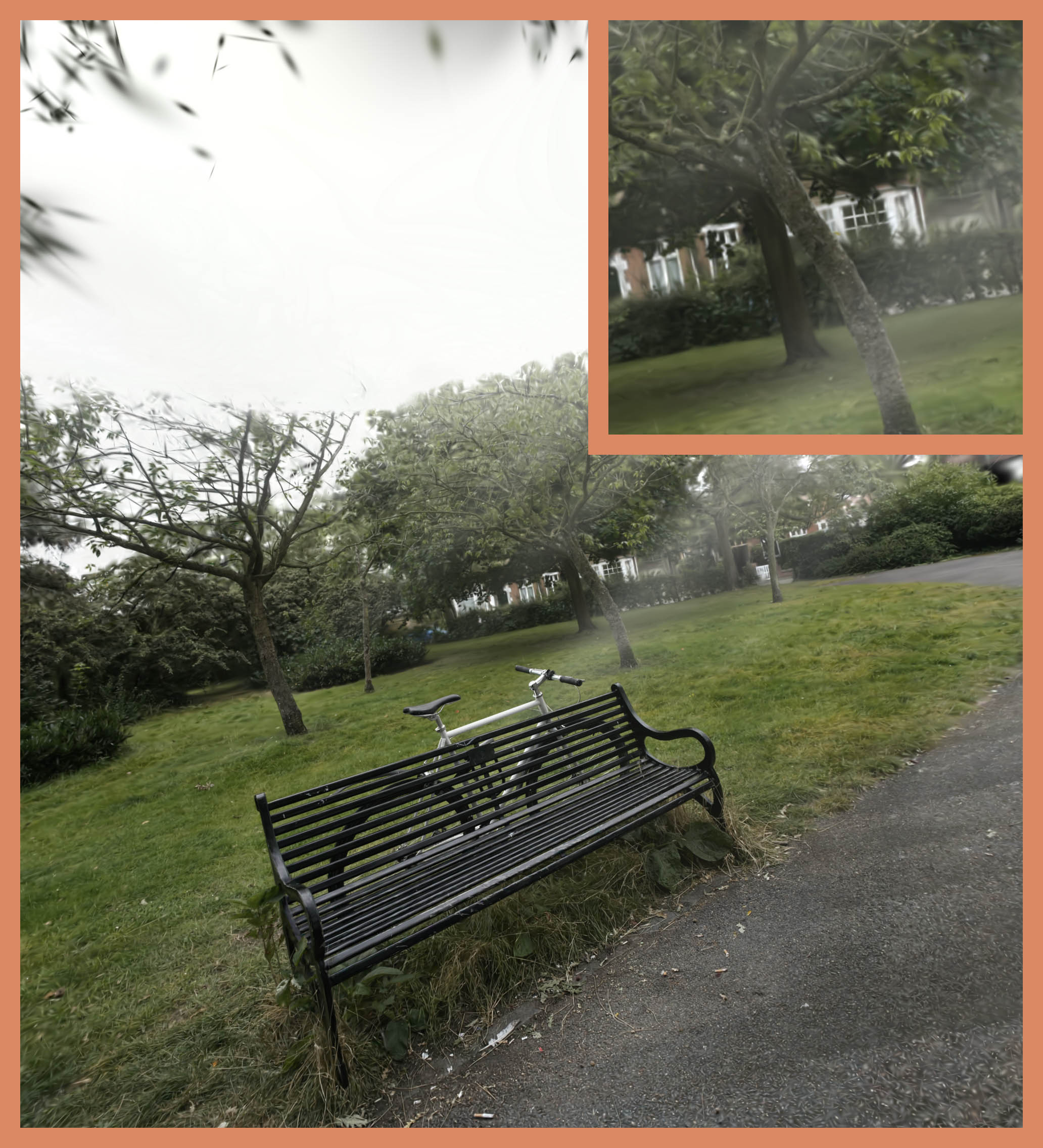}
    \includegraphics[width=0.24\linewidth]{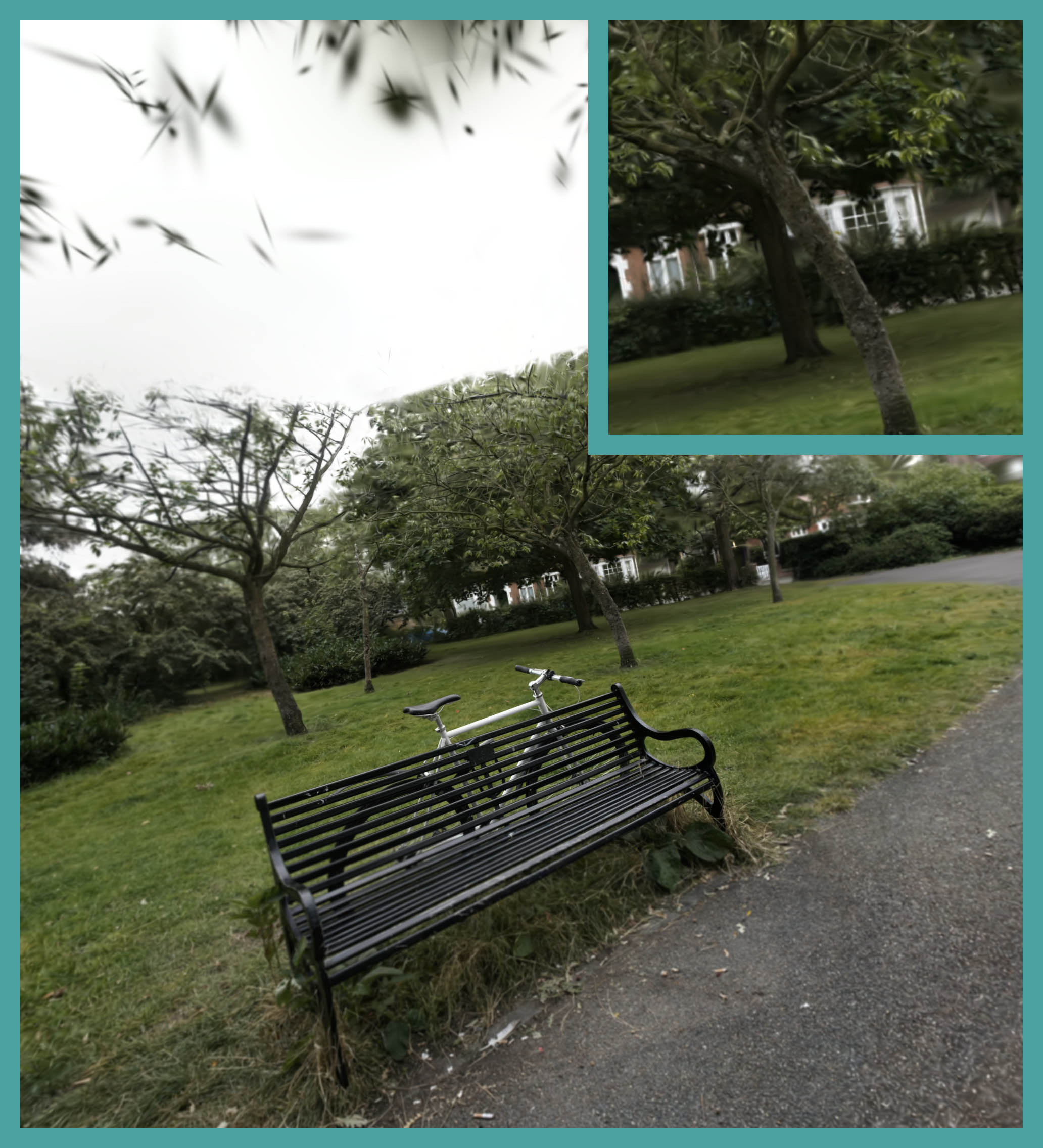}
    \includegraphics[width=0.24\linewidth]{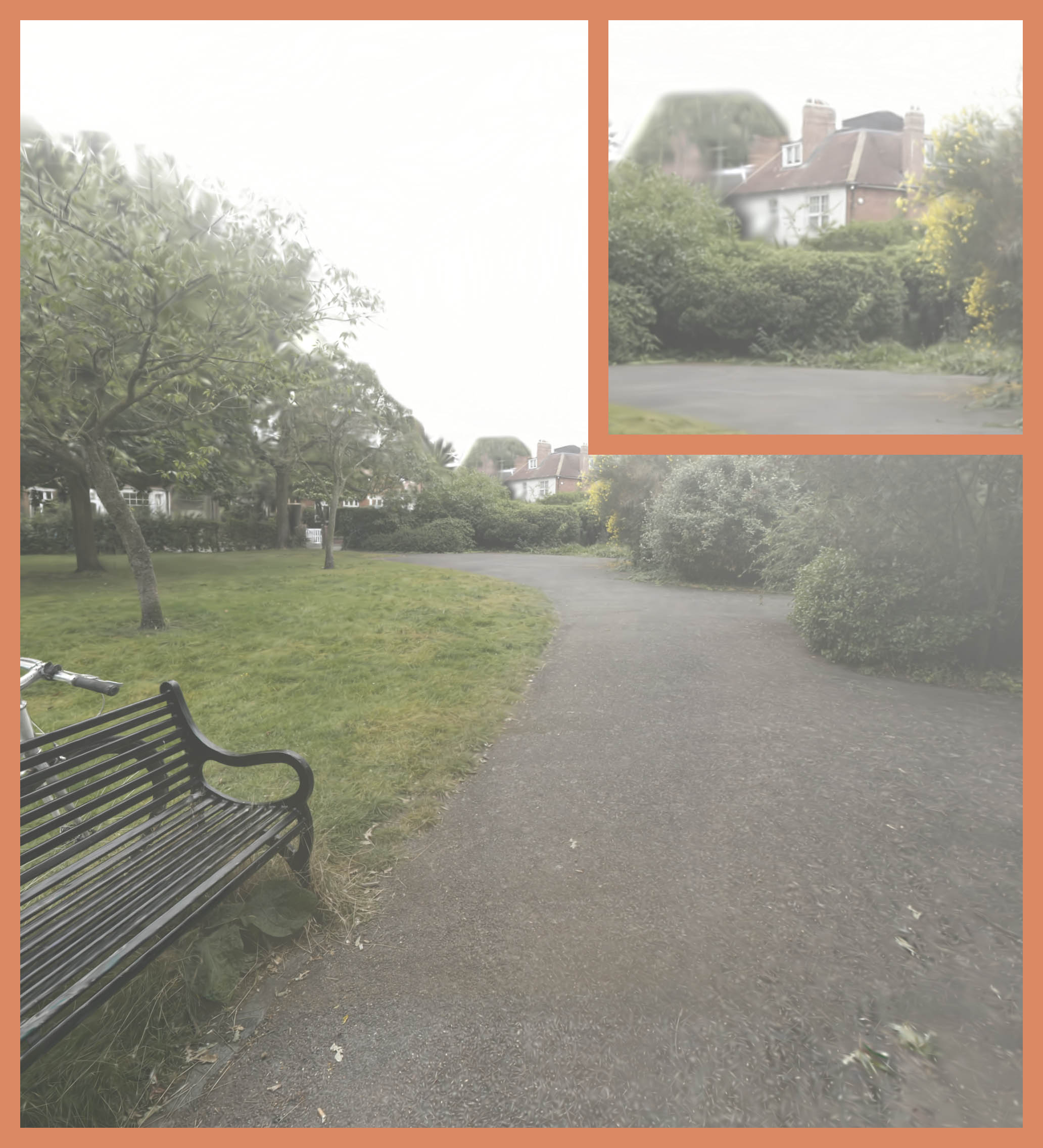}
    \includegraphics[width=0.24\linewidth]{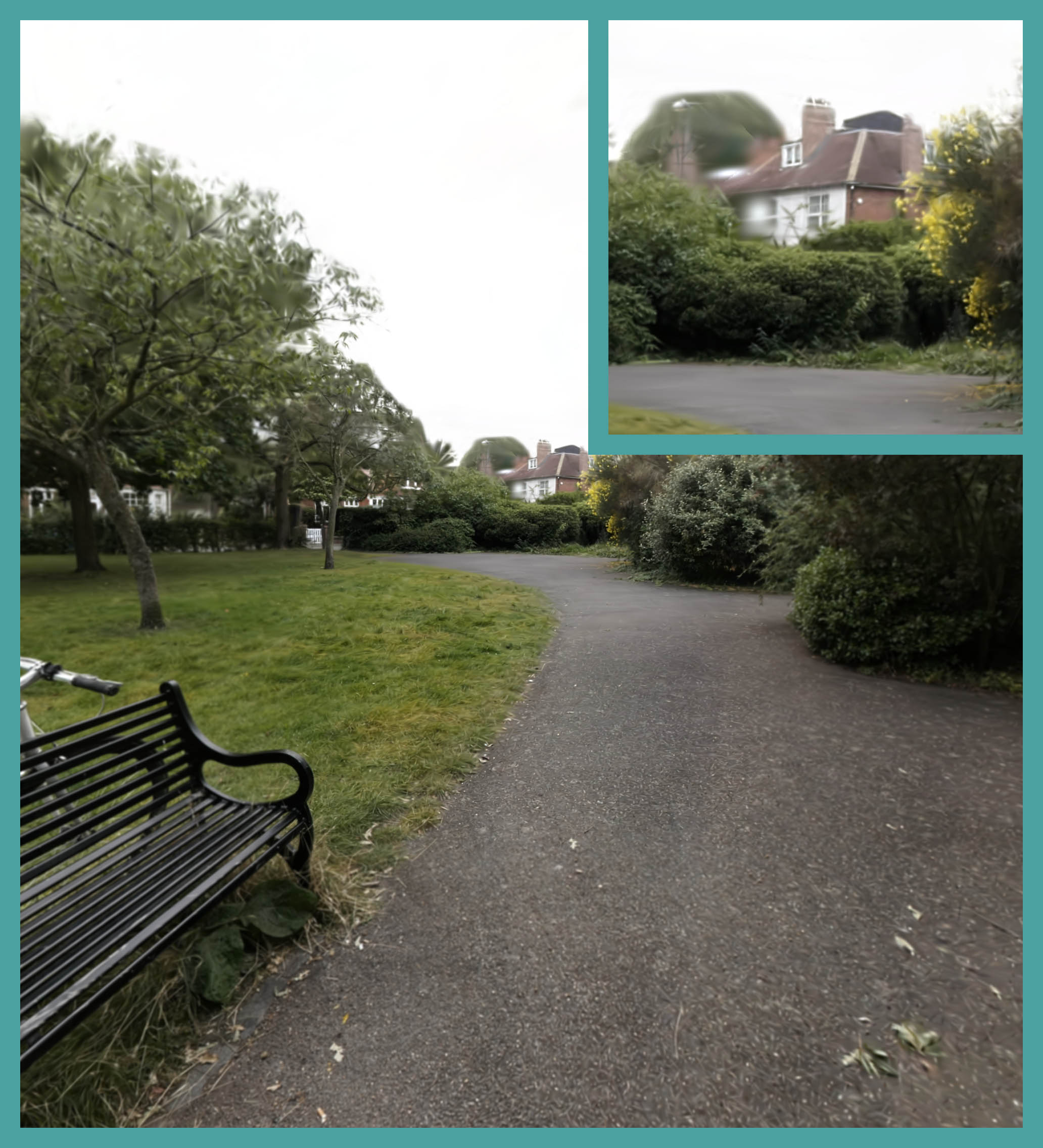}
    \caption{
\new{Normal FOV (top) and VR (bottom) views of \colorbox{secondary_color!70}{Mini-Splatting}~\cite{fang2024minisplatting} and \colorbox{primary_color!70}{ours}.}
Increasing the field-of-view reveals shortcomings of current methods in VR applications:
large Gaussians when projected with the local affine approximation can cause cloud-like artifacts and elongated Gaussians (\ie, \protect\say{god rays}).
\revised{VRSplat eliminates the artifacts while meeting the performance demands of smooth VR experiences.}{VRSplat carefully deals with these issues by employing Optimal Projection~\cite{huang2024optimal}.}}
    \label{fig:artifacts}
\end{figure}

\section{3D Gaussian Splatting}

3DGS~\cite{kerbl3Dgaussians} represents a scene using a mixture of $N$ 3D Gaussians, each given by:
\begin{equation}
    G(\vec x) = e^{-\frac{1}{2}(\vec{x}-\mue)^T\Sigma^{-1}(\vec{x}-\mue)},
\end{equation}
with the Gaussian 3D mean $\mue \in \mathbb{R}^3$, and the 3D covariance matrix $\Sigma = R S S^T R^T$ being constructed from a rotation matrix $R\in \mathbb{R}^{3{\times}3}$ and a diagonal scaling matrix $S\in \mathbb{R}^{3{\times}3}$.
For a pixel $\vec{p}$, the resulting color is given by the volume rendering equation:
\begin{equation}
    {C}(\vec{p}) = \sum_{i=0}^{N} \vec{c}_i \alpha_i \newrevised{\prod_{j=0}^{i-1}}{\sum_{j=0}^{i-1}} (1 - \alpha_j), \label{eq:alphablend}
\end{equation}
where $\alpha_i = \sigma_i G_{\text{2D}}(\vec{p})$ is the alpha value of the projected 2D splat, dependent on the value of the 2D Gaussian $G_{\text{2D}}$ and the opacity value $\sigma_i$, and $\vec{c}_i$ is the Gaussian's view-dependent color.

Rather than considering a fully-sorted list of all primitives for each pixel, 3DGS performs tile-based rasterization;
based on the extent of each Gaussian on the image plane, it is assigned to tiles each covering $16{\times}16$ pixels.
After instantiation of all Gaussian/tile combinations, this arrangement is sorted and  only potentially contributing Gaussians are considered during rendering, simplifying and accelerating the pipeline.

The projection of each 3D Gaussian onto the image plane is demonstrated by~\citet{zwicker_ewa_2001}: 
the covariance matrix of the 2D Gaussian $\sigmatwod$ is obtained by dropping the last row/column of
\begin{equation}
    \Sigma' = J W \Sigma W^T J^T,
\end{equation}
with $W$ the camera matrix and $J$ the local affine approximation of the projective transform.
As noted in~\cite{zwicker_ewa_2001}, this is only correct for the center of the image plane---for Gaussians at extreme positions (\ie on the boundary of the image plane), the approximation error grows larger and causes elongated Gaussians or cloud-like artifacts (see Fig.~\ref{fig:artifacts}).
Optimal Projection~\cite{huang2024optimal} describes an alternative approach to remedy this issue:
Namely, they project Gaussians onto distinct tangent planes on the unit sphere around the camera $\vec{o}$, in turn removing projection-based artifacts with a small performance overhead.

The sort order in Eqn.~\eqref{eq:alphablend} is given by the $z$-coordinate of $\mue$ in view-space.
Clearly, this results in changes of ordering during view rotation, causing popping artifacts~\cite{radl2024stopthepop} --- an artifact also common in other splatting approaches~\cite{mueller1998popping}
\newnew{(\cf the supplemental video).}
While current VR solutions~\cite{cier2024metalsplatter} thus prefer sorting by the distance to the camera, \ie $\|\vec{o} - \mue\|_2$, this now results in popping during translation.
The only viable solution to eliminate all popping artifacts is perform a full sort of primitives considered for each pixel;
despite tile-based rasterization, a full sort is computationally prohibitive if interactive framerates are required.

As the performance of each stage in 3DGS's rendering pipeline is directly tied to primitive counts, reducing the number of Gaussians can drastically improve performance.
While 3DGS's model sizes are tolerable for desktop viewing, the high resolution and demanding framerate requirements of HMDs necessitate smaller Gaussian point clouds.
The underlying root cause of the large primitive counts is 3DGS's heuristics-driven densification~\cite{kerbl3Dgaussians}, which leads to a large number of redundant primitives in over-reconstructed regions.
A plethora of recent works investigate model size reduction whilst retaining high image quality metrics, either through importance-based pruning~\cite{fan2023lightgaussian, niemeyer2024radsplat}, score-based densification~\cite{mallick2024taming}, oversampling and subsequent simplification~\cite{fang2024minisplatting} or sampling-based densification~\cite{kheradmand2024mcmc}.

\section{Method}

\begin{figure}
    \centering
    \includegraphics[width=.95\linewidth]{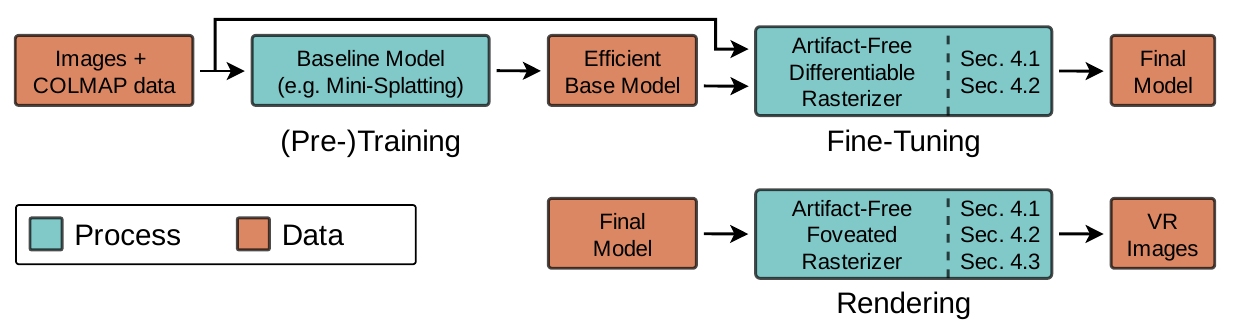}
    \caption{
    \new{An overview of our pipeline for high-fidelity VR rendering:
Starting from an efficient baseline, such as Mini-Splatting~\cite{fang2024minisplatting}, we fine-tune the Gaussians using our artifact-free differentiable rasterizer to arrive at a 3DGS model capable of VR rendering without artifacts.
We thus remain largely baseline-agnostic, and the overhead for our fine-tuning is only a small fraction of the initial baseline reconstruction.
}
    }
    \label{fig:overview}
\end{figure}

This section introduces our full model for immersive VR experiences on consumer-grade HMDs.
We synergize recent advances in 3DGS to arrive at a full model capable of view-consistent rendering of Gaussians.
To address the implied performance penalty, we introduce a novel single-pass foveated rendering solution capable of achieving the required rendering times for immersive VR.
\new{We provide an overview of our complete pipeline in Fig.~\ref{fig:overview}}.

We choose Mini-Splatting~\cite{fang2024minisplatting} as our model backbone, as their more uniform distribution of Gaussians also improves rendering times by balancing per-tile workloads.
Subsequently, we propose a fine-tuning scheme to incorporate StopThePop, Optimal Projection and Mini-Splatting into a single, high-quality representation---which we then accelerate by exploiting foveated rendering.



\subsection{Eliminating Popping Artifacts}
Due to Mini-Splatting representing scenes with a significantly reduced primitive count, Gaussians are naturally larger, and therefore more prone to popping.
Radl and Steiner \emph{et al.}~\cite{radl2024stopthepop} demonstrate that hierarchical per-pixel resorting of Gaussian splats according to depth along view rays is able to overcome popping effectively.
Hence, we opt to use their renderer to achieve view-consistent rendering.

\newnew{3DGS employs its differentiable renderer to optimize Gaussian parameters---including position, rotation, scale, opacity, and spherical harmonics coefficients---via backpropagation.}
Sort order plays an important role during optimization, \ie models need to be rendered with the same sort order as was used during training.
Consequently, na\"ively applying StopThePop to a Mini-Splatting~\cite{fang2024minisplatting} model---which uses a a global sort order during training---reduces image quality.
Therefore, we opt for fine-tuning our small Mini-Splatting models with StopThePop rasterization \revised{(without densification, thus preserving the low primitive count)}{without densification}, which proves to be sufficient to achieve equal or higher scores on standard image quality metrics, \revised{depending}{dependent} on the dataset.
\new{}
This strategy has the additional advantage that it makes our method largely agnostic to the pre-trained model and densification objective, as long as they yield coherent 3D Gaussian representations.

\subsection{Minimizing Projection Error}

The projection-based distortions due to 3DGS's local affine approximation are especially disturbing in VR, where the \revised{large}{wide} FOV and constant movement of the user's head causes cloud-like artifacts to rotate and block the views. 
Optimal Projection for 3DGS~\cite{huang2024optimal} conducted a detailed analysis of the error introduced in the projection step and proposed a solution:
\revised{Namely, each Gaussian is projected to the tangent plane of the unit sphere at $\vec{o}$ that is perpendicular to the line connecting $\vec{o}$ and $\mue$, where $\vec{o}$ denotes the camera position}{Denoting the center of the camera as $\vec{o}$ and the mean of a Gaussian as $\mue$, the Gaussian is projected to the tangent plane of the unit sphere at $\vec{o}$ that is perpendicular to the line connecting $\vec{o}$ and $\mue$}.

\begin{figure}
	\centering
	\includegraphics[width=.7\columnwidth, trim=1.2cm 7.7cm 6cm 2cm, clip]{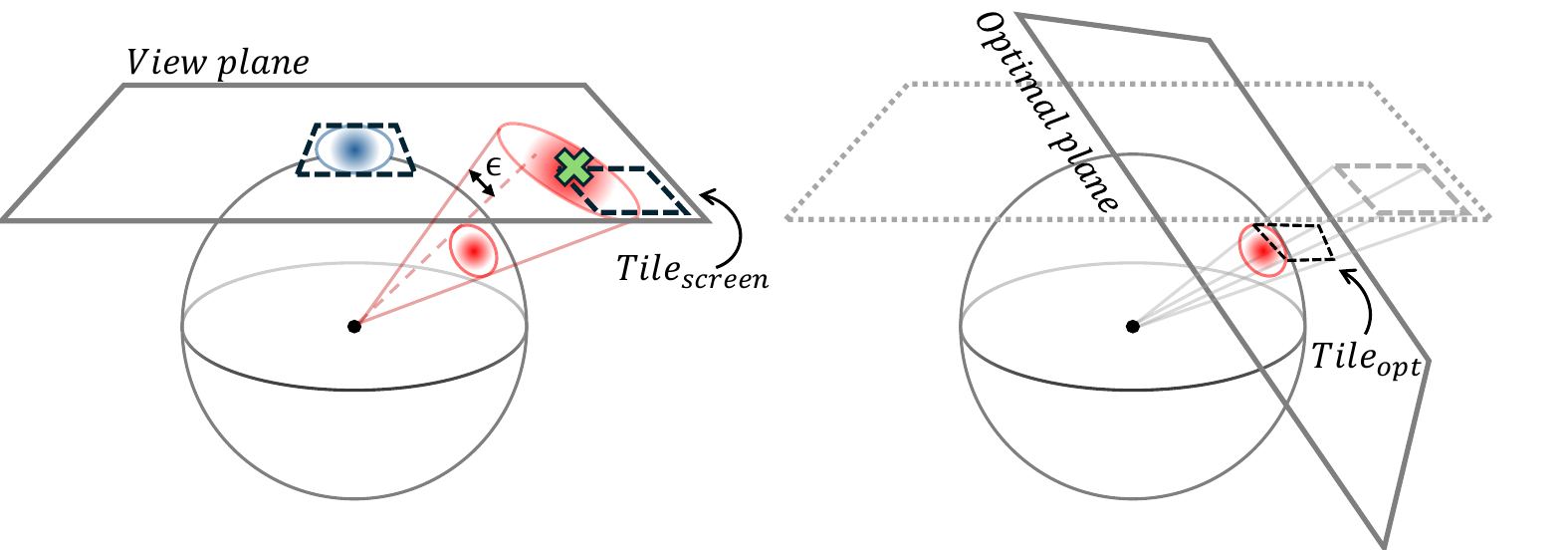} \\
    \begin{minipage}[t]{0.15\linewidth}
        \centering
        \small
        \phantom{StopThePop}
    \end{minipage}
\begin{minipage}[t]{0.34\linewidth}
        \centering
        \small
        \text{StopThePop}
    \end{minipage}
    \begin{minipage}[t]{0.34\linewidth}
        \centering
        \small
        \text{Ours}
    \end{minipage}
\begin{minipage}[t]{0.15\linewidth}
        \centering
        \small
        \phantom{0}
    \end{minipage}
	\caption{
\revised{Left: StopThePop computes the maximum contribution within a tile directly in screen space.
Right: As we employ Optimal Projection~\cite{huang2024optimal}, we find the maximum by reasoning about the maximum on the per-Gaussian optimal plane, leveraging the fact that tiles form general quadrilaterals. Subsequently, projecting this maximum back to screen space yields the maximum contribution within a tile~\cite{anon2024fastandrobust}.}{Due to our use of Optimal Projection~\cite{huang2024optimal}, culling by reasoning about the maximum contribution within a tile is not straightforward.
However, as tiles form quadrilaterals on the optimal plane of each Gaussian, we efficiently compute the maximum by leveraging the fact that the maximum may only lie on at most two edges if the projected mean is not within the tile~\cite{anon2024fastandrobust}.}
    }
	\label{fig:opt}
\end{figure}
However, Optimal Projection is not directly compatible with \revised{tile-based culling, as proposed in}{some of the techniques used} in StopThePop~\cite{radl2024stopthepop}.
\new{Tile-based culling reduces the number of Gaussian/tile combinations by: (1) computing the maximum over $G_{\text{2D}}$ for each Gaussian/tile combination and (2) pruning the Gaussian from the specific tile if $\alpha < \frac{1}{255}$.
Radl and Steiner~\emph{et al.} also show that the maximum over $G_{\text{2D}}$ can be computed by maximizing the Gaussian's contribution over the tile boundaries closest to the 2D mean~\cite{radl2024stopthepop}.
}
\revised{Nevertheless, note that this computation is done fully in screen-space with axis-aligned tiles, thus requiring an adapted algorithm to work with Optimal Projection.
}{
Most prominently, they perform tile-based culling for splats in screen space (\eg, for the red splat in Fig.~\ref{fig:opt}), by finding its maximum density point $\hat{\vec{x}}$ inside an axis-aligned rectangular tile $Tile_{\text{screen}}$, and culling the tile if the density is below a threshold.
In this example, $\hat{\vec{x}}$ is marked using a green cross.}

When the 2D Gaussian's mean $\muetd$ is inside the tile, it follows that \new{the point of maximum contribution} $\hat{\vec{x}}={\muetd}$, otherwise $\hat{\vec{x}}$ must lie on one of the edges that are reachable from $\muetd$.
For Optimal Projection, tile-based culling can no longer be performed on the image plane, but needs to be evaluated on the optimal plane of each Gaussian.
Projecting these screen-space tiles yields again convex quadrilaterals $\newrevised{\mathit{Tile}}{Tile}_{\text{opt}}$.
Consequently, at most two neighboring edges may be reachable from $\muetd$ because projection preserves their geometrical relationship in the original view plane.
For each candidate edge $\vec{p} + t \cdot \vec{d}$, by taking the derivative of density with respect to $t$ and letting it equal zero, the maximum density point along the line can be calculated as
\begin{equation}
\hat{\vec{x}} = \vec{p} + \frac{\vec{d}^\top\sigmatwod^{-1}({\muetd} - \vec{p})}{\vec{d}^\top\sigmatwod^{-1}\vec{d}} \cdot \vec{d},    
\end{equation}
where $\sigmatwod$ is the 2D covariance matrix of the splat.
Evaluating the $\alpha$ value at $\hat{\vec{x}}$ allows us to determine the culling criteria.
Projecting $\hat{\vec{x}}$ back onto the image plane yields the maximum point inside the screen-space tile, which StopThePop requires for its per-tile depth evaluation~\cite{radl2024stopthepop}.
\new{See Fig.~\ref{fig:opt} for a visualization of StopThePop's culling strategy compared to our approach.}

\subsection{Single-Pass Foveated Rendering}
    \label{sec:single_pass_foveated}

To remedy the implied performance penalty of StopThePop rasterization and Optimal Projection, we leverage the reduced visual acuity in the periphery by employing foveated rendering.
While multi-pass solutions (rendering the center and peripheral regions separately) require only minor changes to the pipeline and can be effective when combined with proper culling~\cite{anon2024fastandrobust}, they necessitate processing the same Gaussians multiple times.
Hence, we propose a targeted single-pass solution that ties in elegantly with the tile-based 3DGS rasterizer.

\begin{figure}[h]
    \centering
    \begin{subfigure}[t]{0.45\linewidth}
        \centering
        \includegraphics[width=\linewidth]{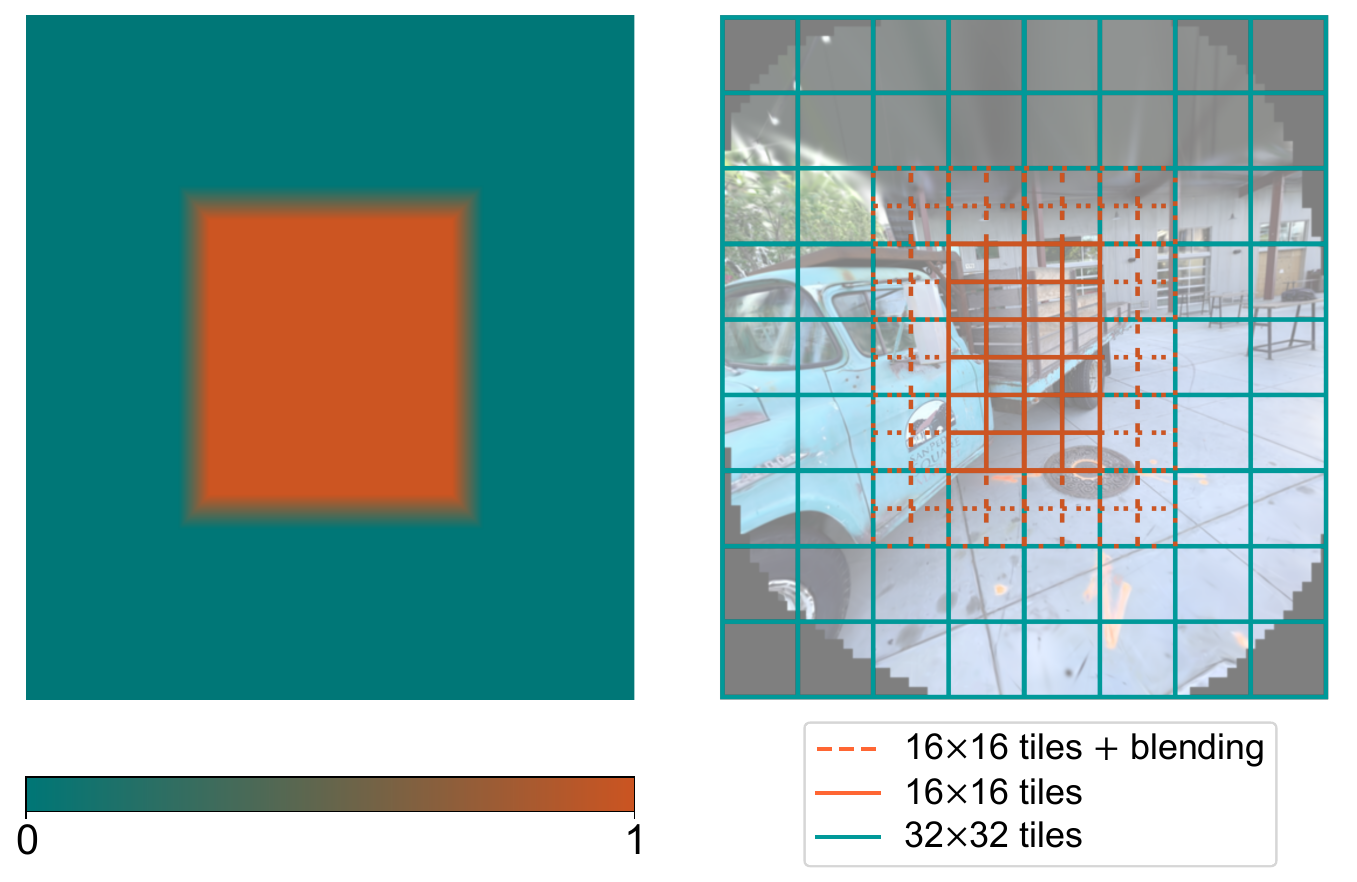}
        \caption{
        Single-pass foveated rendering
        }
        \label{fig:single_pass_foveated}
    \end{subfigure}%
    \hfill
    \begin{subfigure}[t]{0.5\linewidth}
        \centering
        \includegraphics[width=\linewidth]{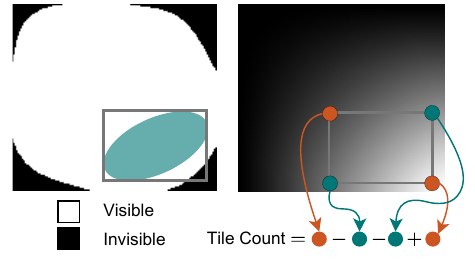}
        \caption{
        Tile visibility mask and Summed Area Table
        }
        \label{fig:visibility_culling}
    \end{subfigure}
    \caption{We design a single-pass foveated rendering routine with visibility culling. (a) Depiction of our chosen tile configurations for our single-pass foveated rendering: 
        we split the center region into tiles of size \colorbox{secondary_color!70}{$16{\times}16$}, and use \colorbox{primary_color!70}{$32{\times}32$} tiles in the periphery; 
        hybrid tiles in-between (dotted) are treated as $16{\times}16$ tiles but the values of groups of $2{\times}2$ pixels are averaged and blended with the individual pixel values, based on the continuous blending mask (left). (b) We sample the visibility mask from OpenXR to first compute a mask of all visible tiles, from which we construct its summed-area table.
        We then compute the exact number of visible tiles covered by the rectangular extent of a 2D Gaussian splat from the summed-area values of its corner points.}
    \label{fig:merged_figure}
\end{figure}

\paragraph{Single-Pass Foveated Rendering.}
In the following, we differentiate only between a high and a low-resolution region by launching differently-sized tiles. 
However, all our strategies can directly be applied to further resolution reductions as well.
As visualized in Fig.~\ref{fig:single_pass_foveated}, we realize the split between high and low resolution regions by using $16{\times}16$ pixel tiles for the center, and $32{\times}32$ tiles for the periphery.
In practice, we always first partition the image into $32{\times}32$ pixel tiles, which are split into four separate subtiles of size $16{\times}16$ in the center region.
This mapping can be efficiently computed at the start of each render pass and enables us to compute the exact number of blocks to launch during the rendering stage.
Consumer-grade \newrevised{HMDs}{HDMs}, like the Meta Quest 3, often do not provide eye tracking, thus we can precompute this mapping once for each eye and re-use it across frames.

During tile-based rasterization, we assign Gaussians to $32{\times}32$ tiles.
This results in fewer Gaussian/tile combinations overall, but additional workload during the rendering stage as subtiles in the center region need to load more Gaussians.
However, we utilize StopThePop's hierarchical culling~\cite{radl2024stopthepop} to mitigate this issue, allowing us to efficiently cull unnecessary Gaussians early for the $16{\times}16$ subtiles.
During the render stage, we launch blocks with $256$ threads (same as \stp and \gs) for both $16{\times}16$ and $32{\times}32$ tiles.
However, each thread in the larger tiles handles groups of $2{\times}2$ pixels and treats them as a single pixel, effectively halving the resolution.


Finally, we blend the high-resolution region with the low-resolution region to avoid noticeable discontinuities.
For the transitional subtiles in the high resolution region (visualized as dotted tiles in Fig.~\ref{fig:single_pass_foveated}), we approximate the color of the low resolution tile by averaging the values of $2{\times}2$ pixels and blend it with the individual pixel values, based on the continuous blending mask.
For the low-resolution region, we perform nearest neighbor upsampling and blurring with a $3{\times}3$ Gaussian kernel, which proves effective in suppressing aliasing artifacts.

\paragraph{Visibility Culling.}

We propose to further accelerate rendering by culling tiles not visible in the HMD.
We process the visibility mask available for the HMD (\eg through OpenXR) to create a bitfield with one bit for each tile, indicating whether or not any pixels of the tile are visible.
\newrevised{Subsequently, we compute a summed-area table from this bitfield, and instantiate Gaussian/tile combinations in two stages:
First, we calculate the exact number of visible tiles each Gaussian may touch using the summed-area table (see Fig.~\ref{fig:visibility_culling});}{Subsequently, we compute a summed-area table from this bitfield, which we then use to compute the exact number of visible tiles a specific Gaussian splat may touch (see Fig.~\ref{fig:visibility_culling}).}
\newnew{We then use this information to allocate the global sorting buffer and to compute the range of each Gaussian's instances inside this buffer.}

This leads to a \newrevised{}{further} reduction in sort entries, as Gaussian/tile combinations outside the visible area of the image are not instantiated.
In our single-pass foveated rendering approach, we can reduce the amount of tiles further by removing the "invisible" tiles entirely from the precomputed mapping, resulting in a \newrevised{}{further} reduction of \revised{${\sim} 11\%$ in the number of tiles and ${\sim}6\%$ in the total number of Gaussian/tile combinations}{${\sim} 6\%$ in the total number of tiles}.


\section{Results and Evaluation}


We use two variants of Mini-Splatting~\cite{fang2024minisplatting} as the base for our model:
the original Mini-Splatting ($z$), where the sort order is given by the view-space $z$-coordinate as well as Mini-Splatting~(Dist), which we modified to order Gaussians according to $\|\mue - \vec{o}\|_2$.
Ours denotes our full method, using a Mini-Splatting backbone fine-tuned for 5K iterations with StopThePop and Optimal Projection.
\new{For both Mini-Splatting and StopThePop, we use the default training parameters.
For fine-tuning, we set the total number of iterations to 35K and start from 30K.
We use tile sizes of $16\times 16$ for center tiles, and $32\times 32$ for peripheral tiles.
The center region is half the rendered image size, with $10\%$ padding of linearly increasing blending weights.
}

We evaluate all methods quantitatively and qualitatively in a controlled user study, using three well-established datasets: the Mip-NeRF 360 dataset~\cite{barron2022mipnerf360}, Tanks \& Temples~\cite{Knapitsch2017tanks} \revised{and}{as well as} Deep Blending~\cite{hedman2018deepblending}.
We perform our performance evaluation and user study using the SIBR framework~\cite{sibr2020bonopera} with OpenXR support and our custom software rasterizer, where we used a Meta Quest 3 tethered to a desktop equipped with an NVIDIA RTX 4090 and render at native resolution ($2064\times 2272$ pixels).
In addition to our method, we also ablated an optimized two-pass foveated renderer to demonstrate the performance improvements of our single-pass approach.
We provide implementation details for this alternative renderer in Appendix~\ref{app:twopassfovea}.

\subsection{User Study}
To enable a well-founded qualitative evaluation of our method, we performed a controlled user study with 25 participants for 3DGS-based VR scene viewing using a Meta Quest 3.
We use Mini-Splatting~\cite{fang2024minisplatting} for our comparison, as 3DGS~\cite{kerbl3Dgaussians} fails to reach the desired framerate target of $\geq$ 72FPS for some scenes due to its excessive primitive count.

We performed two pairwise comparisons:
\begin{itemize}
    \item Ours vs Mini-Splatting ($z$)
    \item Ours vs Mini-Splatting (Dist)
\end{itemize}
Participants were instructed to focus on three distinct criteria: \emph{artifacts}, \emph{quality}, and \emph{preference}, leading to the following hypotheses:
\begin{enumerate}
    \item Our approach produces fewer \emph{artifacts} than other splatting approaches as it avoids popping, elongated Gaussians and cloud-like artifacts.
    \item Using foveated rendering in our approach does not reduce \emph{quality} compared to other approaches.
    \item Our approach is fast and robust while avoiding artifacts and thus leads to a \emph{preferable} VR experience compared to other approaches.
\end{enumerate}

To test these hypotheses, we used a pairwise comparison design \cite{mantiuk2012comparison,kiran2017towards}, where participants experienced a scene rendered with two different methods and then rated them relative to one another. We selected two scenes from the Mip-NeRF 360 dataset~\cite{barron2022mipnerf360} (\emph{Bonsai} and \emph{Bicycle}) and one scene each from Deep Blending~\cite{hedman2018deepblending} and Tanks \& Temples~\cite{Knapitsch2017tanks} (\emph{Dr Johnson} and \emph{Truck}). This selection provided a balanced mix of indoor and outdoor scenes. Each participant viewed two scenes for the Ours vs. Mini-Splatting ($z$) comparison and two for Ours vs. Mini-Splatting (Dist), with the order of scenes and methods randomized to prevent learning effects.

We recruited 25 participants from a local university. After a brief pre-questionnaire, participants rated each pairwise comparison based on general preference, artifacts, and visual quality using a 5-point Likert scale (from \say{Strongly Prefer Method 1} to \say{Strongly Prefer Method 2}), with an option for \say{No Preference.}
Participants were instructed to consider \emph{quality} while standing still and focusing on the central object of the current scene, and to move around frequently to assess potential \emph{artifacts}. Each participant could freely explore the scene but could not switch back and forth between conditions. While there was no strict time limit, we encouraged them to transition to the next condition after approximately 90 seconds. Responses were converted into scores $s \in \{-2, -1, 0, +1, +2\}$, where $-2$ indicated a strong preference for Mini-Splatting and $+2$ indicated a strong preference for our method.


\begin{figure}[ht!]
	\centering
	\includegraphics[width=.9\linewidth]{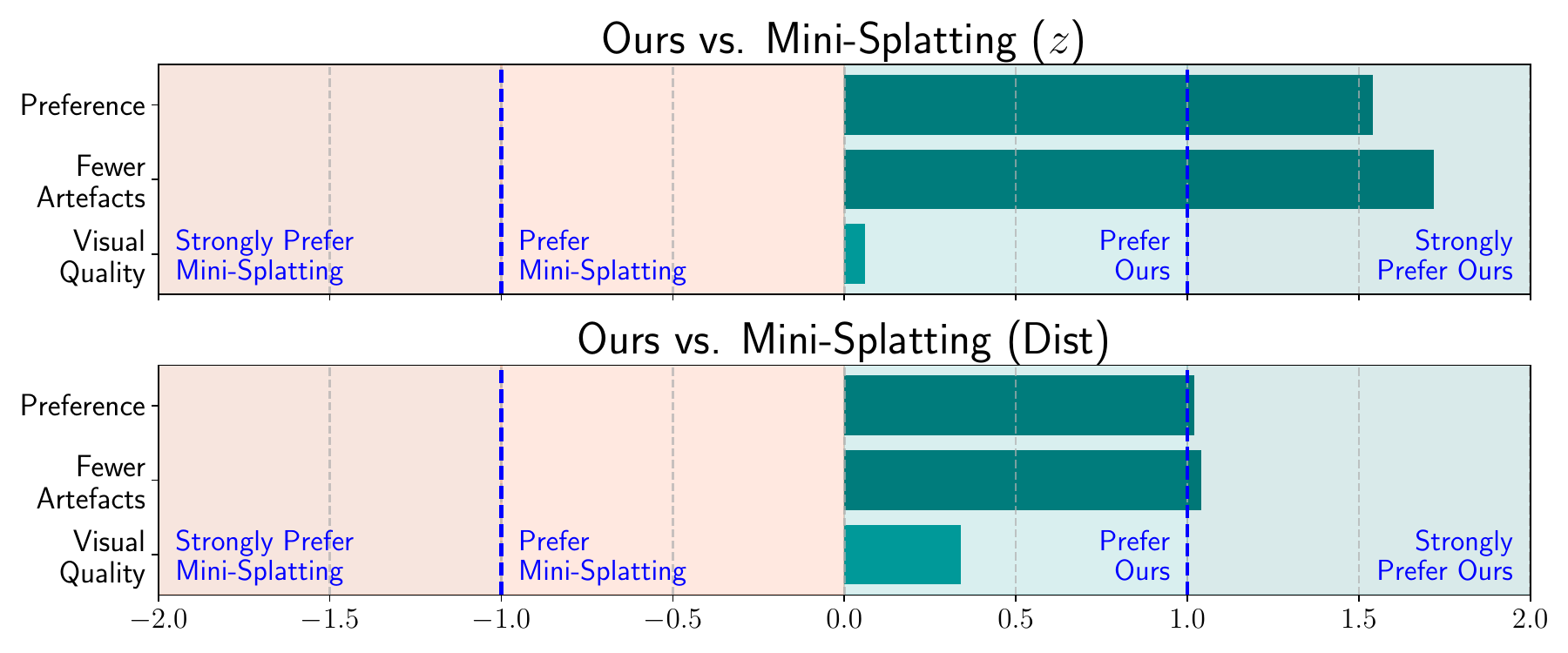}
	\caption{
User Study results: 
our method is preferred by users across all tested modalities including quality, artifacts and preference.
Due to the prevalence of popping artifacts in Mini-Splatting ($z$), users strongly prefer our method.
In Mini-Splatting (Dist), projection errors cause the preference for our method.
}
	\label{fig:userstudy}
\end{figure}

To analyze the results, we conducted Wilcoxon T-tests on the pairwise results. 
The average results are shown in Fig.~\ref{fig:userstudy}.
For the Ours vs. Mini-Splatting ($z$) comparison, both \emph{preference} ($t=18.0$, $p<.001$) and \emph{artifacts} ($t=0.0$, $p<.001$) were statistically significant, while \emph{quality} ($t=209.5$, $p=.62$) was not.
For the Ours vs. Mini-Splatting (Dist) comparison, all measures were statistically significant: \emph{preference} ($t=61.0$, $p<.001$), \emph{artifacts} ($t=15.0$, $p<.001$) and \emph{quality} ($t=121.0$, $p<.03$).


Clearly, our method is preferred by a large margin in terms of general preference and artifacts.
Mini-Splatting ($z$) was rated poorly due to immersion-breaking popping artifacts and errors due to the projection.
In contrast to quantitative evaluation on still images, Mini-Splatting obtains \emph{better} qualitative ratings in VR when sorting by the Euclidean distance instead. However, due to projection errors and occasional popping artifacts, Ours is still preferred by a significant margin. Please refer to the supplemental video for a direct visual comparison and illustration of artifacts.
In terms of visual quality, our method is preferred by a small margin despite our foveated rendering solution---this shows that with our blending CUDA kernel, we can effectively reduce resolution in the peripheral region without concerns about visual quality.
Although we did not directly compare Mini-Splatting ($z$) and Mini-Splatting (Dist), the comparison with our method suggests that sorting by Euclidean distance reduces popping artifacts to some extent, \revised{slightly improving the VR experience, while the same projection-based artifacts remain disturbing.}{but at the cost of reconstruction quality, as reflected in the quality metrics in Tab.~\ref{tab:image_metrics_large_fov}}. 

In summary, all of our hypotheses were supported, as confirmed by participant feedback. Popping and non-uniform artifacts are significant issues for Mini-Splatting in VR, which our method avoids entirely through the combination of StopThePop and Optimal Projection. Participants commented: 
\say{The lighting changes abruptly when I move!} for Mini-Splatting and \say{This is so much more stable} for our approach. Our combination of strategies, including foveated rendering, does not reduce quality and even outperforms Mini-Splatting (Dist) in terms of user-perceived quality.
One participant noted for Mini-Splatting:  \say{When I stand still there are still some unnatural God rays here. And they are even different for the left and right eye.} For our method: \say{This is clearly better, with no weird artifacts, and it stays consistent when I move my head. Both methods seem equally fast.}
While we did not specifically compare our method to 3DGS, previous work suggests that the same preference trends prevail~\cite{anon2024fastandrobust}.
For more details about the user study and per-scene results, see Appendix~\ref{app:userstudy}.

\subsection{Image Metrics}
We compare our final method against 3DGS~\cite{kerbl3Dgaussians}, Mini-Splatting~\cite{fang2024minisplatting}, and StopThePop~\cite{radl2024stopthepop} in terms of visual quality.
We use PSNR, SSIM, LPIPS~\cite{zhang2018unreasonable}\revised{}{and {\FLIP}~\cite{Andersson2020Flip}} and present our results in Tab.~\ref{tab:image_metrics_large_fov}.

\begin{figure}
    \centering
    \includegraphics[width=\linewidth]{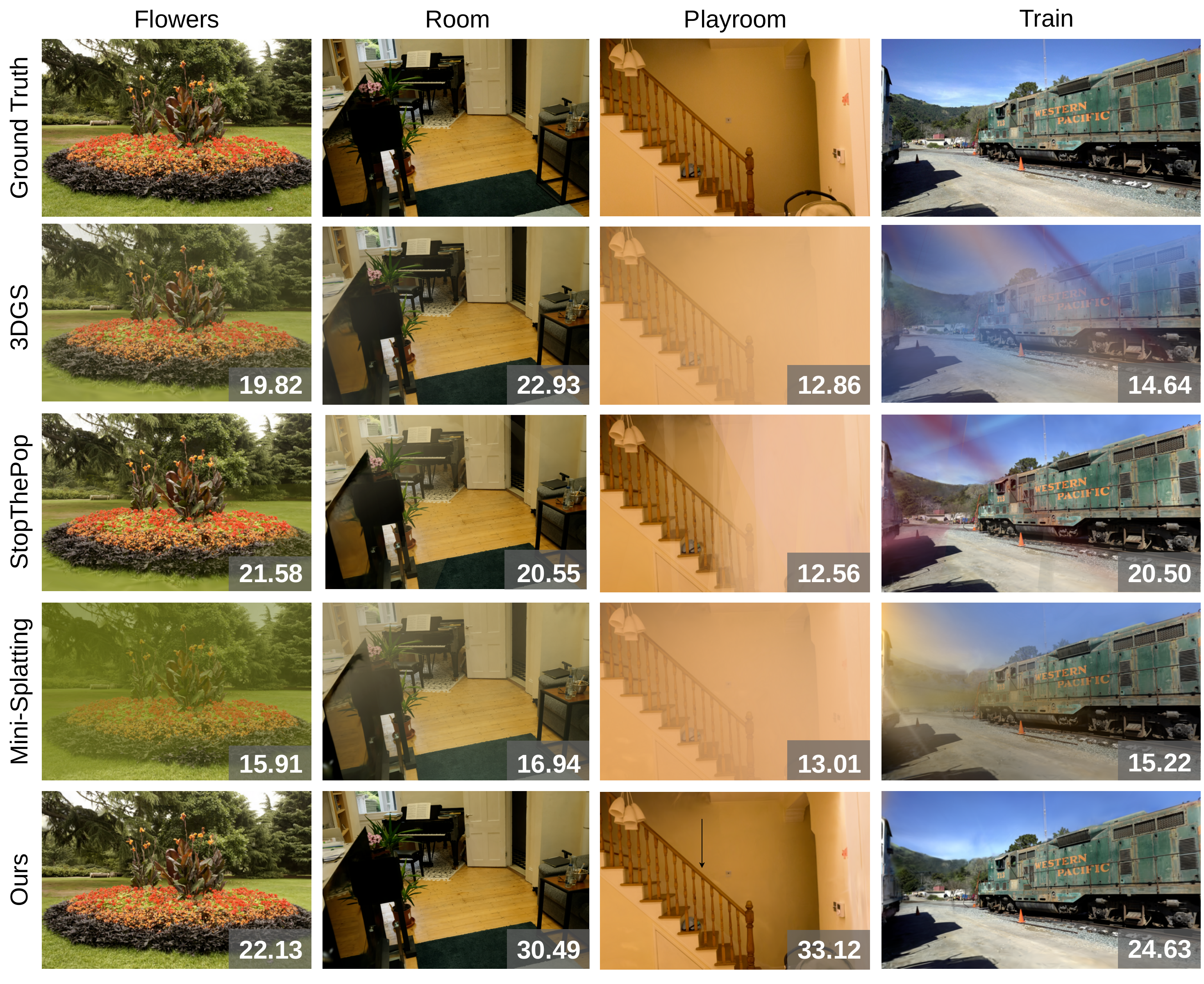}
    \caption{\new{Images rendered using our large FOV evaluation protocol with inset PSNR values. 
Projection errors in other methods lead to disturbing artifacts, as also observed in a VR setting.}}
    \label{fig:large_fov_collage}
\end{figure}

\new{Similar to~\citet{huang2024optimal}, we follow an adapted protocol, which highlights our robustness to projection errors for large FOVs, as used in VR. 
Namely, we decrease focal length and increase the render resolution by a factor of $3\times$, which produces elongated and cloud-like Gaussians with standard 3DGS rasterization.
We then use a center crop, allowing for a pixel-perfect cutout that can be compared against the test images.
Without projection errors, this large FOV cutout should exactly match the low FOV render.
As can be seen in Tab.~\ref{tab:image_metrics_large_fov}, our approach consistently beats related work in this scenario.
We also show some qualitative comparisons in Fig.~\ref{fig:large_fov_collage}.}

\new{We also provide image metrics following the standard setup Appendix~\ref{app:image_metrics}}. 
When comparing the numbers across tables, our artifact-free rasterizer is able to retain the image quality, whereas competing methods suffer from severely reduced image metrics.
Overall, this evaluation setup demonstrates our improved image quality when considering the requirements for VR rendering.


\begin{table*}[ht!]
    \centering
    \caption{
\new{Standard image quality metrics for our large FOV evaluation protocol.
Our proposed method does not suffer from projection artifacts and, therefore, outperforms all other methods on a variety of datasets.}
}\setlength{\tabcolsep}{5pt}
    \footnotesize
\begin{tabular}{lrrrrrrrrr}
\toprule
Dataset & \multicolumn{3}{c}{Mip-NeRF 360} & \multicolumn{3}{c}{Tanks \& Temples} & \multicolumn{3}{c}{Deep Blending}\\
\cmidrule(lr){2-4} \cmidrule(lr){5-7} \cmidrule(lr){8-10}
 & PSNR\textsuperscript{$\uparrow$} & SSIM\textsuperscript{$\uparrow$} & LPIPS\textsuperscript{$\downarrow$} & PSNR\textsuperscript{$\uparrow$} & SSIM\textsuperscript{$\uparrow$} & LPIPS\textsuperscript{$\downarrow$} & PSNR\textsuperscript{$\uparrow$} & SSIM\textsuperscript{$\uparrow$} & LPIPS\textsuperscript{$\downarrow$}  \\
\midrule
StopThePop & \cellcolor{tab_color!10} 27.040 & \cellcolor{tab_color!10} 0.812 & \cellcolor{tab_color!30} 0.213 & \cellcolor{tab_color!10} 20.241 & \cellcolor{tab_color!10} 0.809 & \cellcolor{tab_color!50} 0.192 & \cellcolor{tab_color!10} 27.553 & \cellcolor{tab_color!10} 0.889 & \cellcolor{tab_color!10} 0.243 \\
3DGS & 26.820 & 0.804 & 0.219 & 17.112 & 0.741 & 0.229 & 26.192 & 0.875 & 0.247 \\
\midrule
Mini-Splatting ($z$) & 24.765 & 0.786 & 0.229 & 15.355 & 0.687 & 0.284 & 24.347 & 0.871 & 0.262 \\
Mini-Splatting (Dist) & 25.373 & 0.794 & 0.225 & 17.018 & 0.754 & 0.239 & 26.133 & 0.880 & 0.255 \\
Ours ($z$) & \cellcolor{tab_color!50} 27.293 & \cellcolor{tab_color!50} 0.823 & \cellcolor{tab_color!50} 0.212 & \cellcolor{tab_color!50} 22.901 & \cellcolor{tab_color!50} 0.836 & \cellcolor{tab_color!30} 0.197 & \cellcolor{tab_color!30} 30.333 & \cellcolor{tab_color!30} 0.912 & \cellcolor{tab_color!30} 0.242 \\
Ours (Dist) & \cellcolor{tab_color!30} 27.262 & \cellcolor{tab_color!30} 0.822 & \cellcolor{tab_color!10} 0.213 & \cellcolor{tab_color!30} 22.840 & \cellcolor{tab_color!30} 0.835 & \cellcolor{tab_color!10} 0.198 & \cellcolor{tab_color!50} 30.337 & \cellcolor{tab_color!50} 0.912 & \cellcolor{tab_color!50} 0.242 \\
\bottomrule
\end{tabular}
    
    \label{tab:image_metrics_large_fov}
\end{table*}

\subsection{Ablation Studies}

\begin{figure}
    \centering
    \includegraphics[width=.8\linewidth]{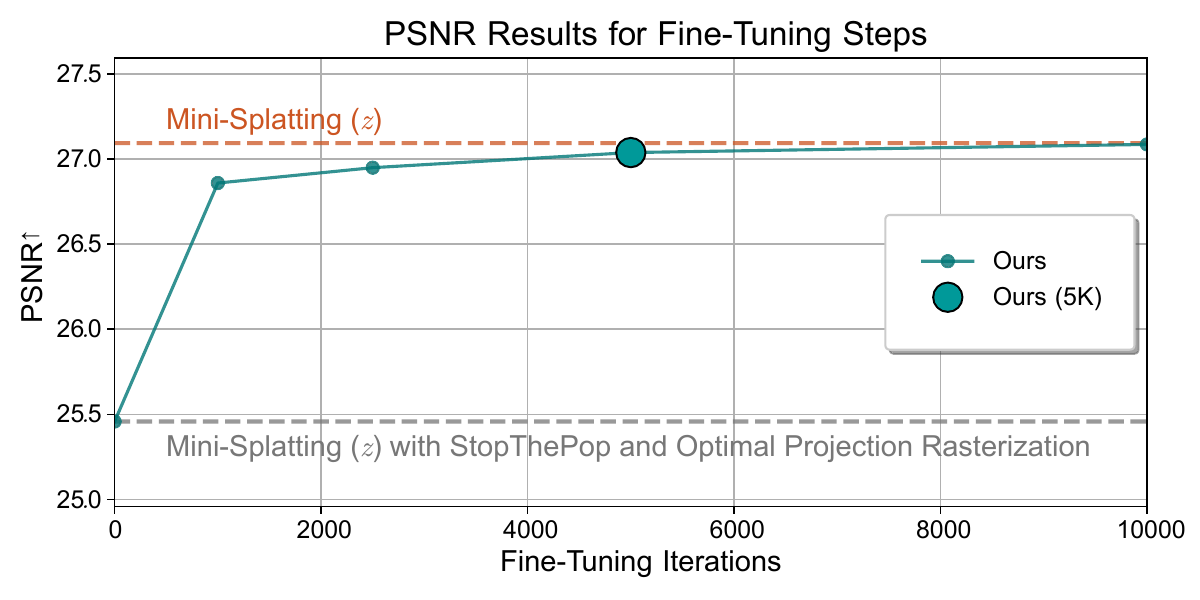}
    \caption{
Ablation study on the number of \revised{}{StopThePop~\cite{radl2024stopthepop}} fine-tuning iterations.
We show PSNR averaged over all tested scenes.
Whereas the initial quality is significantly reduced with \revised{the new rasterization}{hierarchical rasterization}, metrics recover quickly and approach Mini-Splatting ($z$)~\cite{fang2024minisplatting} with 5K fine-tuning iterations.
}
    \label{fig:ablation_train}
\end{figure}

To validate our design choice of \revised{fine-tuning with StopThePop~\cite{radl2024stopthepop} and Optimal Projection~\cite{huang2024optimal} rasterization}{StopThePop~\cite{radl2024stopthepop} fine-tuning} for 5K iterations, we run a small-scale ablation study.
We evaluate Mini-Splatting ($z$)~\cite{fang2024minisplatting} \revised{after $\{1\text{K},\ 2.5\text{K},\ 5\text{K},\ 10\text{K}\}$ iterations of fine-tuning}{with hierarchically resorted rendering after $\{1\text{K},\ 2.5\text{K},\ 5\text{K},\ 10\text{K}\}$ iterations} and plot the PSNR in Fig.~\ref{fig:ablation_train}. 
We include the PSNR score of Mini-Splatting \revised{as well as}{in addition to} Mini-Splatting with \revised{the new rasterization}{hierarchical resorting} (no fine-tuning) as baselines.

\begin{figure}
    \centering
    \includegraphics[width=\linewidth]{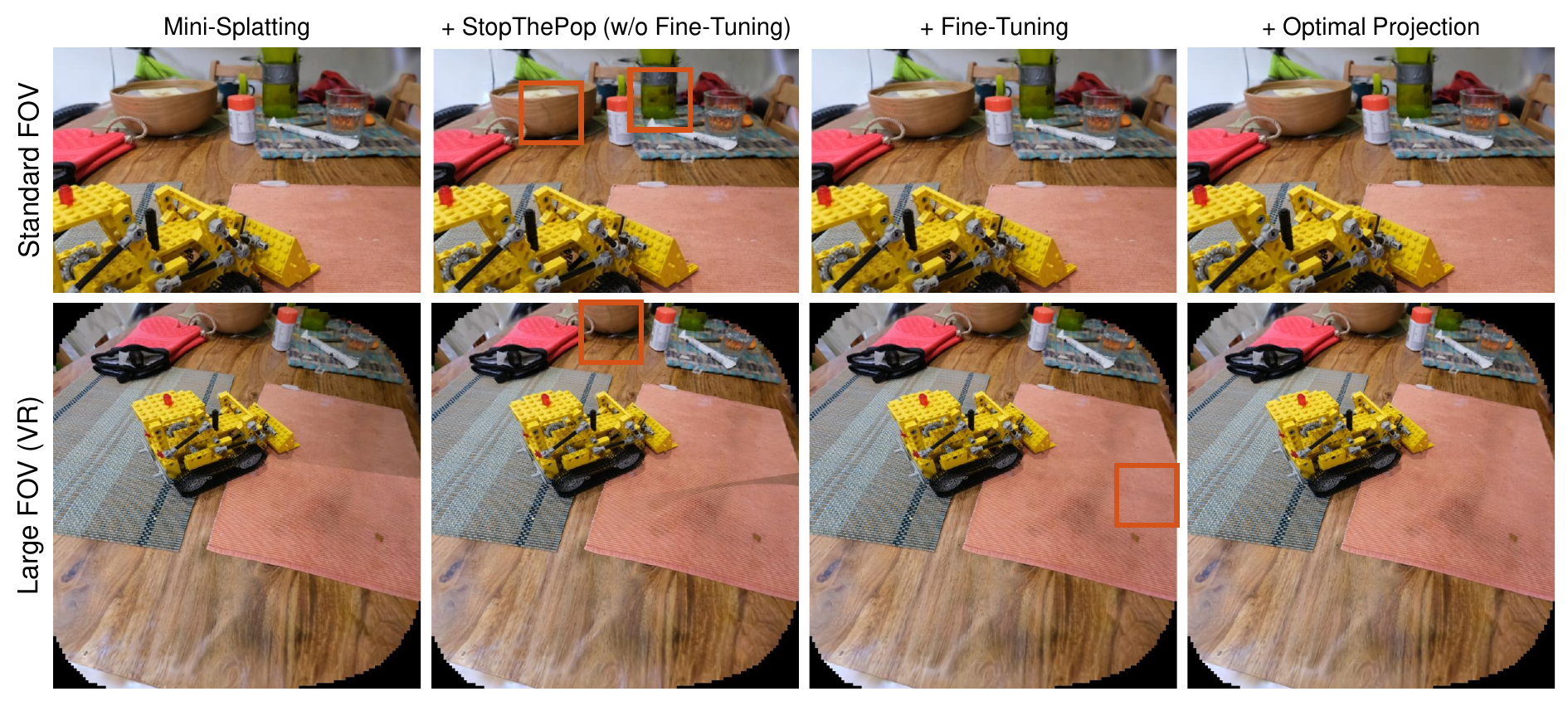}
    \caption{
\newnew{Qualitative evaluation of our method's components.
When rendering a Mini-Splatting~\cite{fang2024minisplatting} model directly with StopThePop, the changes in sort-order lead to artifacts for both normal and large FOV.
Our fine-tuning step allows the model to learn this updated sort order.
Finally, Optimal Projection removes artifacts due to projection errors, when rendering with a larger FOV}
}
    \label{fig:ablation_images}
\end{figure}

Mini-Splatting with \revised{the new renderer}{StopThePop rendering} initially decreases image quality as the ordering of Gaussians diverges from the learned ordering, causing vastly different per-pixel RGB colors and a significant drop in image quality (${\sim} 1.5$ dB). 
However, image quality recovers quickly and already approaches the results of Mini-Splatting after 5K fine-tuning steps.
Since $5\text{K}$ additional training iterations only result in a minor performance improvement ($27.04$ vs. $27.09$ dB), we choose $5\text{K}$ fine-tuning iterations in the interest of fast optimization.
See \revised{Appendix~\ref{sec:app:finetuning}}{our provided supplementary material} for detailed results and a discussion.
\newnew{We also provide a qualitative evaluation of our method's components in Fig.~\ref{fig:ablation_images}.}

\subsection{Performance Timings}

We provide additional render timings to quantify the performance impact of the different techniques in our pipeline.
The images were rendered in the resolution as requested by the Meta Quest 3, \ie $2064{\times}2272$, and timings were doubled to recreate the stereoscopic setting.
To obtain representative timings, we interpolate a camera path between all provided training and test set cameras for each scene and average our measurements over $4$ runs.
We use CUDA 12.6 with an NVIDIA RTX 4090 for all presented measurements.

The general performance target for a smooth VR experience on the Meta Quest 3 is 72 FPS (${\sim}14$ms).
As we show in Tab.~\ref{tab:performance}, \gs is able to reach this target despite its large model sizes.
\minigs's considerably smaller models can lead to large performance gains of up to $2.2{\times}$.
Adding \stp on top of \minigs results in a large increase in rendering times due to the additional sorting overhead.
Note that all these methods suffer either from popping and/or projection error artifacts.

\begin{table}[]
    \centering
    
    \caption{
Average performance timings (in ms) per dataset with our VR evaluation setup (two images with $2064{\times}2272$ resolution).
Our average render times are below the $14$ms threshold, and significantly outperform the simple two-pass foveated rendering routine.
}
    \small
    \setlength{\tabcolsep}{5pt}
    \begin{tabular}{lccccc}
    \toprule
    \multirow{2}{*}{Timings in ms} && \multirow{2}{*}{Deep Blending} & \multicolumn{2}{c}{Mip-NeRF 360} & \multirow{2}{*}{Tanks \& Temples} \\\cmidrule(lr){4-5}
     && & Indoor & Outdoor & \\
    \midrule
    \gs && \cellcolor{tab_color!10} 8.65 & \cellcolor{tab_color!30} 8.06 & \cellcolor{tab_color!30} 10.35 & \cellcolor{tab_color!10} 9.94 \\
    \minigs && \cellcolor{tab_color!50} 4.51 & \cellcolor{tab_color!50} 6.24 & \cellcolor{tab_color!50} 4.98 & \cellcolor{tab_color!50} 4.76 \\
    \minigs + \stp && \cellcolor{tab_color!0} 10.55 & \cellcolor{tab_color!0} 14.48 & \cellcolor{tab_color!0} 11.11 & \cellcolor{tab_color!0} 10.95 \\
    \ours (w/o fov.) && \cellcolor{tab_color!0} 14.54 & \cellcolor{tab_color!0} 20.48 & \cellcolor{tab_color!0} 15.91 & \cellcolor{tab_color!0} 15.42 \\
    \ours (two-pass fov.) && \cellcolor{tab_color!0} 10.35 &\cellcolor{tab_color!0}  15.01 & \cellcolor{tab_color!0} 12.49 & \cellcolor{tab_color!0} 11.18 \\
    \ours && \cellcolor{tab_color!30} 8.14 & \cellcolor{tab_color!10} 12.36 & \cellcolor{tab_color!10} 10.47 & \cellcolor{tab_color!30} 9.22 \\
    \bottomrule
    \end{tabular}

    \label{tab:performance}
\end{table}

\ours without any foveated rendering (w/o fov.) is slower than \minigs + \stp due to the additional overhead from Optimal Projection.
A simple two-pass foveated rendering approach accelerates our method, however, there are still inefficiencies due to a duplicate rendering of the blended region (transitional band) and redundant processing of Gaussians in the early, per-Gaussian stages of the rasterizer.

\ours allows us to fully utilize synergies, by fusing kernel calls and rendering the blended region only once.
Additionally, as we assign Gaussians to large $32{\times}32$ tiles, the workload during Gaussian/tile instantiation is significantly reduced, while the implied rendering overhead is largely mitigated by StopThePop's hierarchical culling.
Overall, \ours reaches framerates comparable to \gs and clearly stays below the $14$ms render budget for most scenes and viewpoints, while providing an artifact-free VR experience.

\new{The image resolution of $2064\times 2272$ results in a total of 4615 tiles (of size $32\times 32$), which are further divided into 1085 high-resolution tiles, 2921 low-resolution tiles, 102 hybrid tiles (where blending is performed), and 507 invisible tiles.}

We investigate per-scene performance timings in more detail in Tab.~\ref{tab:performance_per_scene}, where we see that performance gains between \gs and \minigs correlate heavily with model size (number of Gaussians).
For example, \emph{Truck} sees a massive decrease in model size by $12{\times}$ and performance improvements of $2.11{\times}$, while \emph{Counter}'s model size only decreases by $3{\times}$ and performance improves by $1.15{\times}$.
In scenes, where primitive counts cannot be sufficiently decreased (e.g. \emph{Bonsai}), \ours is unable to beat \gs's performance, mostly due to the per-pixel sorting overhead.
Note that performance is also influenced by factors other than model size, \revised{such as}{like} Gaussian extent, opacity values, and distribution of Gaussians in the scene.

\begin{table}[]
    \centering

    \caption{Average performance timings (in ms) per scene for all datasets with our VR evaluation setup (two images with $2064{\times}2272$ resolution). 
    Additionally, we report the number of Gaussians $N$ (in millions) for \gs and \minigs/\ours.
    Our full method outperforms the two-pass foveated solution and stays within the predefined render budget.
}
    \scriptsize
    \setlength{\tabcolsep}{1.5pt}
    
    \begin{tabular}{lccccccccccccc}
        \toprule
        Scene & Bicycle & Flowers & Garden & Stump & Treehill & Bonsai & Counter & Kitchen & Room & DrJ & Playroom & Train & Truck\\\cmidrule(lr){2-14}
        $N$ (\gs) & 5.95M & 3.60M & 5.49M & 4.84M & 3.85M & 1.25M & 1.20M & 1.81M & 1.55M & 3.28M & 2.33M & 1.05M & 2.56M \\
        $N$ (\minigs/\ours) & 0.52M & 0.56M & 0.56M & 1.00M & 0.55M & 0.36M & 0.41M & 0.43M & 0.43M & 0.37M & 0.32M & 0.19M & 0.21M \\
        \midrule
        \gs & \cellcolor{tab_color!0} 13.97 & \cellcolor{tab_color!30} 8.53 & \cellcolor{tab_color!30} 11.91 & \cellcolor{tab_color!30} 8.45 & \cellcolor{tab_color!30} 8.90 &
        \cellcolor{tab_color!30} 7.32 & \cellcolor{tab_color!30} 8.22 & \cellcolor{tab_color!30} 9.46 & \cellcolor{tab_color!30} 7.24 & \cellcolor{tab_color!10} 9.39 & \cellcolor{tab_color!30} 7.91 & \cellcolor{tab_color!10} 9.81 & \cellcolor{tab_color!10} 10.06 \\
        \minigs (MS) & \cellcolor{tab_color!50} 4.91 & \cellcolor{tab_color!50} 4.99 & \cellcolor{tab_color!50} 5.06 & \cellcolor{tab_color!50} 4.99 & \cellcolor{tab_color!50} 4.97 & \cellcolor{tab_color!50} 6.26 & \cellcolor{tab_color!50} 7.11 & \cellcolor{tab_color!50} 6.31 & \cellcolor{tab_color!50} 5.26 & \cellcolor{tab_color!50} 4.69 & \cellcolor{tab_color!50} 4.34 & \cellcolor{tab_color!50} 4.76 & \cellcolor{tab_color!50} 4.76 \\
        MS + \stp& \cellcolor{tab_color!10} 11.12 & \cellcolor{tab_color!0} 10.37 & \cellcolor{tab_color!10} 12.18 & \cellcolor{tab_color!0} 10.91 & \cellcolor{tab_color!0} 10.97 & \cellcolor{tab_color!0} 14.39 & \cellcolor{tab_color!0} 16.72 & \cellcolor{tab_color!0} 14.89 & \cellcolor{tab_color!0} 11.91 & \cellcolor{tab_color!0} 10.55 & \cellcolor{tab_color!0} 10.54 & \cellcolor{tab_color!0} 10.73 & \cellcolor{tab_color!0} 11.18\\
        \ours (w/o fov.) & \cellcolor{tab_color!0} 15.92 & \cellcolor{tab_color!0} 14.91 & \cellcolor{tab_color!0} 17.84 & \cellcolor{tab_color!0} 15.17 & \cellcolor{tab_color!0} 15.71 & \cellcolor{tab_color!0} 20.26 & \cellcolor{tab_color!0} 23.65 & \cellcolor{tab_color!0} 21.22 & \cellcolor{tab_color!0} 16.80 & \cellcolor{tab_color!0} 14.22 & \cellcolor{tab_color!0} 14.86 & \cellcolor{tab_color!0} 15.01 & \cellcolor{tab_color!0} 15.83\\
        \ours (two-pass fov.) & \cellcolor{tab_color!0} 12.39 & \cellcolor{tab_color!0} 11.83 & \cellcolor{tab_color!0} 14.24 & \cellcolor{tab_color!0} 11.70 & \cellcolor{tab_color!0} 12.28 & \cellcolor{tab_color!0} 15.24 & \cellcolor{tab_color!0} 16.87 & \cellcolor{tab_color!0} 15.56 & \cellcolor{tab_color!0} 12.36 & \cellcolor{tab_color!0} 10.15 & \cellcolor{tab_color!0} 10.55 & \cellcolor{tab_color!0} 10.74 & \cellcolor{tab_color!0} 11.63\\
        \ours & \cellcolor{tab_color!30} 10.14 & \cellcolor{tab_color!10} 9.89 & \cellcolor{tab_color!0} 12.19 & \cellcolor{tab_color!10} 9.80 & \cellcolor{tab_color!10} 10.35 & \cellcolor{tab_color!10} 12.60 & \cellcolor{tab_color!10} 14.11 & \cellcolor{tab_color!10} 12.97 & \cellcolor{tab_color!10} 9.76 & \cellcolor{tab_color!30} 7.88 & \cellcolor{tab_color!10} 8.39 & \cellcolor{tab_color!30} 8.89 & \cellcolor{tab_color!30} 9.55\\
        \bottomrule

    \end{tabular}

    \label{tab:performance_per_scene}
\end{table}

\section{Conclusion, Limitations and Future Work}

In this work, we identified the limitations of 3D Gaussian Splatting in delivering high-quality virtual reality experiences. 
By reviewing recent advances that address some of these challenges, we developed an elegant solution that integrates multiple techniques to enable immersive, artifact-free VR rendering. 
In addition, we presented a novel single-pass foveated rendering solution building on the hierarchical StopThePop rasterizer.
Our implementation consistently achieves real-time framerates across all tested scenes on a Meta Quest 3 HMD. 
Furthermore, our formal user study validates these results, demonstrating a clear preference for our method across a diverse set of scenes.

Although our approach successfully mitigates the most prominent artifacts, approximate hierarchical depth sorting may still cause flickering in regions with complex geometric relationships. 
This highlights the need for a robust level-of-detail (LOD) scheme or fully accurate volume rendering for 3D Gaussians. 
While ray-tracing 3D Gaussians is a potential solution, recent studies~\cite{moenneloccoz2024gaussianraytracing, blanc2024raygauss} indicate that it currently remains too computationally expensive for high-quality VR experiences.
There are also areas for optimization in our VR rendering pipeline, such as monocular rendering and stereoscopic reuse of distant content, and the fusion of stereo rendering passes to maximize synergy effects.

The code for fine-tuning our models and the VR viewer is publicly available at {\color{blue}\url{https://github.com/Cekavis/VRSplat}}.

\begin{acks}
LR, MS, \& MS: This research was funded in whole or in part by the Austrian Science Fund (FWF) [10.55776/I6663]. For open access purposes, the author has applied a CC BY public copyright license to any author-accepted manuscript version arising from this submission.
\end{acks}

\bibliographystyle{ACM-Reference-Format}
\bibliography{i3d_revised}     


\begin{thebibliography}{46}


\ifx \showCODEN    \undefined \def \showCODEN     #1{\unskip}     \fi
\ifx \showDOI      \undefined \def \showDOI       #1{#1}\fi
\ifx \showISBNx    \undefined \def \showISBNx     #1{\unskip}     \fi
\ifx \showISBNxiii \undefined \def \showISBNxiii  #1{\unskip}     \fi
\ifx \showISSN     \undefined \def \showISSN      #1{\unskip}     \fi
\ifx \showLCCN     \undefined \def \showLCCN      #1{\unskip}     \fi
\ifx \shownote     \undefined \def \shownote      #1{#1}          \fi
\ifx \showarticletitle \undefined \def \showarticletitle #1{#1}   \fi
\ifx \showURL      \undefined \def \showURL       {\relax}        \fi
\providecommand\bibfield[2]{#2}
\providecommand\bibinfo[2]{#2}
\providecommand\natexlab[1]{#1}
\providecommand\showeprint[2][]{arXiv:#2}

\bibitem[Barron et~al\mbox{.}(2022)]%
        {barron2022mipnerf360}
\bibfield{author}{\bibinfo{person}{Jonathan~T. Barron}, \bibinfo{person}{Ben Mildenhall}, \bibinfo{person}{Dor Verbin}, \bibinfo{person}{Pratul~P. Srinivasan}, {and} \bibinfo{person}{Peter Hedman}.} \bibinfo{year}{2022}\natexlab{}.
\newblock \showarticletitle{{Mip-NeRF 360: Unbounded Anti-Aliased Neural Radiance Fields}}. In \bibinfo{booktitle}{\emph{IEEE/CVF Conference on Computer Vision and Pattern Recognition}}.
\newblock


\bibitem[Barron et~al\mbox{.}(2023)]%
        {barron2023zip}
\bibfield{author}{\bibinfo{person}{Jonathan~T. Barron}, \bibinfo{person}{Ben Mildenhall}, \bibinfo{person}{Dor Verbin}, \bibinfo{person}{Pratul~P. Srinivasan}, {and} \bibinfo{person}{Peter Hedman}.} \bibinfo{year}{2023}\natexlab{}.
\newblock \showarticletitle{{Zip-NeRF: Anti-Aliased Grid-Based Neural Radiance Fields}}. In \bibinfo{booktitle}{\emph{IEEE/CVF International Conference on Computer Vision}}.
\newblock


\bibitem[Blanc et~al\mbox{.}(2024)]%
        {blanc2024raygauss}
\bibfield{author}{\bibinfo{person}{Hugo Blanc}, \bibinfo{person}{Jean-Emmanuel Deschaud}, {and} \bibinfo{person}{Alexis Paljic}.} \bibinfo{year}{2024}\natexlab{}.
\newblock \bibinfo{title}{{RayGauss: Volumetric Gaussian-Based Ray Casting for Photorealistic Novel View Synthesis}}.
\newblock
\newblock
\showeprint[arxiv]{2408.03356}~[cs.CV]
\urldef\tempurl%
\url{https://arxiv.org/abs/2408.03356}
\showURL{%
\tempurl}


\bibitem[Bonopera et~al\mbox{.}(2020)]%
        {sibr2020bonopera}
\bibfield{author}{\bibinfo{person}{Sebastien Bonopera}, \bibinfo{person}{Jerome Esnault}, \bibinfo{person}{Siddhant Prakash}, \bibinfo{person}{Simon Rodriguez}, \bibinfo{person}{Theo Thonat}, \bibinfo{person}{Mehdi Benadel}, \bibinfo{person}{Gaurav Chaurasia}, \bibinfo{person}{Julien Philip}, {and} \bibinfo{person}{George Drettakis}.} \bibinfo{year}{2020}\natexlab{}.
\newblock \bibinfo{title}{sibr: A System for Image Based Rendering}.
\newblock
\newblock
\urldef\tempurl%
\url{https://gitlab.inria.fr/sibr/sibr_core}
\showURL{%
\tempurl}


\bibitem[Chakravarthula et~al\mbox{.}(2021)]%
        {chakravarthula2021gaze}
\bibfield{author}{\bibinfo{person}{Praneeth Chakravarthula}, \bibinfo{person}{Zhan Zhang}, \bibinfo{person}{Okan Tursun}, \bibinfo{person}{Piotr Didyk}, \bibinfo{person}{Qi Sun}, {and} \bibinfo{person}{Henry Fuchs}.} \bibinfo{year}{2021}\natexlab{}.
\newblock \showarticletitle{Gaze-contingent retinal speckle suppression for perceptually-matched foveated holographic displays}.
\newblock \bibinfo{journal}{\emph{IEEE Transactions on Visualization and Computer Graphics}} \bibinfo{volume}{27}, \bibinfo{number}{11} (\bibinfo{year}{2021}), \bibinfo{pages}{4194--4203}.
\newblock


\bibitem[Cier et~al\mbox{.}(2024)]%
        {cier2024metalsplatter}
\bibfield{author}{\bibinfo{person}{Sean Cier}, \bibinfo{person}{Xijie Yang}, \bibinfo{person}{Anton Maridi}, {and} \bibinfo{person}{Kem Chen}.} \bibinfo{year}{2024}\natexlab{}.
\newblock \bibinfo{title}{{MetalSplatter}}.
\newblock
\newblock
\urldef\tempurl%
\url{https://github.com/scier/MetalSplatter}
\showURL{%
\tempurl}


\bibitem[Deng et~al\mbox{.}(2022)]%
        {deng2022fovnerf}
\bibfield{author}{\bibinfo{person}{Nianchen Deng}, \bibinfo{person}{Zhenyi He}, \bibinfo{person}{Jiannan Ye}, \bibinfo{person}{Budmonde Duinkharjav}, \bibinfo{person}{Praneeth Chakravarthula}, \bibinfo{person}{Xubo Yang}, {and} \bibinfo{person}{Qi Sun}.} \bibinfo{year}{2022}\natexlab{}.
\newblock \showarticletitle{{FoV}-{NeRF}: {Foveated} {Neural} {Radiance} {Fields} for {Virtual} {Reality}}.
\newblock \bibinfo{journal}{\emph{IEEE Transactions on Visualization and Computer Graphics}} \bibinfo{volume}{28}, \bibinfo{number}{11} (\bibinfo{year}{2022}), \bibinfo{pages}{3854--3864}.
\newblock


\bibitem[Fan et~al\mbox{.}(2024)]%
        {fan2023lightgaussian}
\bibfield{author}{\bibinfo{person}{Zhiwen Fan}, \bibinfo{person}{Kevin Wang}, \bibinfo{person}{Kairun Wen}, \bibinfo{person}{Zehao Zhu}, \bibinfo{person}{Dejia Xu}, {and} \bibinfo{person}{Zhangyang Wang}.} \bibinfo{year}{2024}\natexlab{}.
\newblock \showarticletitle{{LightGaussian: Unbounded 3D Gaussian Compression with 15x Reduction and 200+ FPS}}. In \bibinfo{booktitle}{\emph{Advances in Neural Information Processing Systems}}.
\newblock


\bibitem[Fang and Wang(2024)]%
        {fang2024minisplatting}
\bibfield{author}{\bibinfo{person}{Guangchi Fang} {and} \bibinfo{person}{Bing Wang}.} \bibinfo{year}{2024}\natexlab{}.
\newblock \showarticletitle{{Mini-Splatting: Representing Scenes with a Constrained Number of Gaussians}}. In \bibinfo{booktitle}{\emph{European Conference on Computer Vision}}.
\newblock


\bibitem[Fridovich-Keil et~al\mbox{.}(2022)]%
        {fridovich2022plenoxels}
\bibfield{author}{\bibinfo{person}{Sara Fridovich-Keil}, \bibinfo{person}{Alex Yu}, \bibinfo{person}{Matthew Tancik}, \bibinfo{person}{Qinhong Chen}, \bibinfo{person}{Benjamin Recht}, {and} \bibinfo{person}{Angjoo Kanazawa}.} \bibinfo{year}{2022}\natexlab{}.
\newblock \showarticletitle{{Plenoxels: Radiance Fields without Neural Networks}}. In \bibinfo{booktitle}{\emph{IEEE/CVF Conference on Computer Vision and Pattern Recognition}}.
\newblock


\bibitem[Guenter et~al\mbox{.}(2012)]%
        {guenter2012foveated}
\bibfield{author}{\bibinfo{person}{Brian Guenter}, \bibinfo{person}{Mark Finch}, \bibinfo{person}{Steven Drucker}, \bibinfo{person}{Desney Tan}, {and} \bibinfo{person}{John Snyder}.} \bibinfo{year}{2012}\natexlab{}.
\newblock \showarticletitle{{Foveated 3D Graphics}}.
\newblock \bibinfo{journal}{\emph{ACM TOG}} \bibinfo{volume}{31}, \bibinfo{number}{6}, Article \bibinfo{articleno}{164} (\bibinfo{year}{2012}), \bibinfo{numpages}{10}~pages.
\newblock


\bibitem[Hedman et~al\mbox{.}(2018)]%
        {hedman2018deepblending}
\bibfield{author}{\bibinfo{person}{Peter Hedman}, \bibinfo{person}{Julien Philip}, \bibinfo{person}{True Price}, \bibinfo{person}{Jan-Michael Frahm}, \bibinfo{person}{George Drettakis}, {and} \bibinfo{person}{Gabriel Brostow}.} \bibinfo{year}{2018}\natexlab{}.
\newblock \showarticletitle{{Deep Blending for Free-viewpoint Image-based Rendering}}.
\newblock \bibinfo{journal}{\emph{ACM TOG}} \bibinfo{volume}{37}, \bibinfo{number}{6}, Article \bibinfo{articleno}{257} (\bibinfo{year}{2018}), \bibinfo{numpages}{15}~pages.
\newblock


\bibitem[Huang et~al\mbox{.}(2024)]%
        {huang2024optimal}
\bibfield{author}{\bibinfo{person}{Letian Huang}, \bibinfo{person}{Jiayang Bai}, \bibinfo{person}{Jie Guo}, \bibinfo{person}{Yuanqi Li}, {and} \bibinfo{person}{Yanwen Guo}.} \bibinfo{year}{2024}\natexlab{}.
\newblock \showarticletitle{{On the Error Analysis of 3D Gaussian Splatting and an Optimal Projection Strategy}}. In \bibinfo{booktitle}{\emph{European Conference on Computer Vision}}.
\newblock


\bibitem[Jiang et~al\mbox{.}(2024)]%
        {jiang2024vrgs}
\bibfield{author}{\bibinfo{person}{Ying Jiang}, \bibinfo{person}{Chang Yu}, \bibinfo{person}{Tianyi Xie}, \bibinfo{person}{Xuan Li}, \bibinfo{person}{Yutao Feng}, \bibinfo{person}{Huamin Wang}, \bibinfo{person}{Minchen Li}, \bibinfo{person}{Henry Lau}, \bibinfo{person}{Feng Gao}, \bibinfo{person}{Yin Yang}, {and} \bibinfo{person}{Chenfanfu Jiang}.} \bibinfo{year}{2024}\natexlab{}.
\newblock \showarticletitle{{VR-GS: A Physical Dynamics-Aware Interactive Gaussian Splatting System in Virtual Reality}}. In \bibinfo{booktitle}{\emph{SIGGRAPH Conference Papers}}. Article \bibinfo{articleno}{78}, \bibinfo{numpages}{1}~pages.
\newblock


\bibitem[Kaplanyan et~al\mbox{.}(2019)]%
        {kaplanyan2019deepfovea}
\bibfield{author}{\bibinfo{person}{Anton~S Kaplanyan}, \bibinfo{person}{Anton Sochenov}, \bibinfo{person}{Thomas Leimk{\"u}hler}, \bibinfo{person}{Mikhail Okunev}, \bibinfo{person}{Todd Goodall}, {and} \bibinfo{person}{Gizem Rufo}.} \bibinfo{year}{2019}\natexlab{}.
\newblock \showarticletitle{{DeepFovea: Neural Reconstruction for Foveated Rendering and Video Compression using Learned Statistics of Natural Videos}}.
\newblock \bibinfo{journal}{\emph{ACM TOG}} \bibinfo{volume}{38}, \bibinfo{number}{6} (\bibinfo{year}{2019}), \bibinfo{pages}{1--13}.
\newblock


\bibitem[Kerbl et~al\mbox{.}(2023)]%
        {kerbl3Dgaussians}
\bibfield{author}{\bibinfo{person}{Bernhard Kerbl}, \bibinfo{person}{Georgios Kopanas}, \bibinfo{person}{Thomas Leimk{\"u}hler}, {and} \bibinfo{person}{George Drettakis}.} \bibinfo{year}{2023}\natexlab{}.
\newblock \showarticletitle{{3D Gaussian Splatting for Real-Time Radiance Field Rendering}}.
\newblock \bibinfo{journal}{\emph{ACM TOG}} \bibinfo{volume}{42}, \bibinfo{number}{4} (\bibinfo{year}{2023}).
\newblock


\bibitem[Kerbl et~al\mbox{.}(2024)]%
        {kerbl2024hierarchical}
\bibfield{author}{\bibinfo{person}{Bernhard Kerbl}, \bibinfo{person}{Andreas Meuleman}, \bibinfo{person}{Georgios Kopanas}, \bibinfo{person}{Michael Wimmer}, \bibinfo{person}{Alexandre Lanvin}, {and} \bibinfo{person}{George Drettakis}.} \bibinfo{year}{2024}\natexlab{}.
\newblock \showarticletitle{{A Hierarchical 3D Gaussian Representation for Real-Time Rendering of Very Large Datasets}}.
\newblock \bibinfo{journal}{\emph{ACM TOG}} \bibinfo{volume}{43}, \bibinfo{number}{4} (\bibinfo{year}{2024}).
\newblock


\bibitem[Kheradmand et~al\mbox{.}(2024)]%
        {kheradmand2024mcmc}
\bibfield{author}{\bibinfo{person}{Shakiba Kheradmand}, \bibinfo{person}{Daniel Rebain}, \bibinfo{person}{Gopal Sharma}, \bibinfo{person}{Weiwei Sun}, \bibinfo{person}{Yang-Che Tseng}, \bibinfo{person}{Hossam Isack}, \bibinfo{person}{Abhishek Kar}, \bibinfo{person}{Andrea Tagliasacchi}, {and} \bibinfo{person}{Kwang~Moo Yi}.} \bibinfo{year}{2024}\natexlab{}.
\newblock \showarticletitle{{3D Gaussian Splatting as Markov Chain Monte Carlo}}. In \bibinfo{booktitle}{\emph{Advances in Neural Information Processing Systems}}.
\newblock


\bibitem[Kiran~Adhikarla et~al\mbox{.}(2017)]%
        {kiran2017towards}
\bibfield{author}{\bibinfo{person}{Vamsi Kiran~Adhikarla}, \bibinfo{person}{Marek Vinkler}, \bibinfo{person}{Denis Sumin}, \bibinfo{person}{Rafal~K Mantiuk}, \bibinfo{person}{Karol Myszkowski}, \bibinfo{person}{Hans-Peter Seidel}, {and} \bibinfo{person}{Piotr Didyk}.} \bibinfo{year}{2017}\natexlab{}.
\newblock \showarticletitle{Towards a quality metric for dense light fields}. In \bibinfo{booktitle}{\emph{IEEE/CVF Conference on Computer Vision and Pattern Recognition}}.
\newblock


\bibitem[Knapitsch et~al\mbox{.}(2017)]%
        {Knapitsch2017tanks}
\bibfield{author}{\bibinfo{person}{Arno Knapitsch}, \bibinfo{person}{Jaesik Park}, \bibinfo{person}{Qian-Yi Zhou}, {and} \bibinfo{person}{Vladlen Koltun}.} \bibinfo{year}{2017}\natexlab{}.
\newblock \showarticletitle{{Tanks and Temples: Benchmarking Large-Scale Scene Reconstruction}}.
\newblock \bibinfo{journal}{\emph{ACM TOG}} \bibinfo{volume}{36}, \bibinfo{number}{4}, Article \bibinfo{articleno}{78} (\bibinfo{year}{2017}), \bibinfo{numpages}{13}~pages.
\newblock


\bibitem[Konrad et~al\mbox{.}(2020)]%
        {konrad2020gaze}
\bibfield{author}{\bibinfo{person}{Robert Konrad}, \bibinfo{person}{Anastasios Angelopoulos}, {and} \bibinfo{person}{Gordon Wetzstein}.} \bibinfo{year}{2020}\natexlab{}.
\newblock \showarticletitle{Gaze-Contingent Ocular Parallax Rendering for Virtual Reality}.
\newblock \bibinfo{journal}{\emph{ACM TOG}} \bibinfo{volume}{39}, \bibinfo{number}{2} (\bibinfo{year}{2020}), \bibinfo{pages}{1--12}.
\newblock


\bibitem[Krajancich et~al\mbox{.}(2023)]%
        {krajancich2023towards}
\bibfield{author}{\bibinfo{person}{Brooke Krajancich}, \bibinfo{person}{Petr Kellnhofer}, {and} \bibinfo{person}{Gordon Wetzstein}.} \bibinfo{year}{2023}\natexlab{}.
\newblock \showarticletitle{Towards Attention-aware Foveated Rendering}.
\newblock \bibinfo{journal}{\emph{ACM TOG}} \bibinfo{volume}{42}, \bibinfo{number}{4} (\bibinfo{year}{2023}), \bibinfo{pages}{1--10}.
\newblock


\bibitem[Lin et~al\mbox{.}(2024b)]%
        {lin2024vastgaussian}
\bibfield{author}{\bibinfo{person}{Jiaqi Lin}, \bibinfo{person}{Zhihao Li}, \bibinfo{person}{Xiao Tang}, \bibinfo{person}{Jianzhuang Liu}, \bibinfo{person}{Shiyong Liu}, \bibinfo{person}{Jiayue Liu}, \bibinfo{person}{Yangdi Lu}, \bibinfo{person}{Xiaofei Wu}, \bibinfo{person}{Songcen Xu}, \bibinfo{person}{Youliang Yan}, {and} \bibinfo{person}{Wenming Yang}.} \bibinfo{year}{2024}\natexlab{b}.
\newblock \showarticletitle{{VastGaussian: Vast 3D Gaussians for Large Scene Reconstruction}}. In \bibinfo{booktitle}{\emph{IEEE/CVF Conference on Computer Vision and Pattern Recognition}}.
\newblock


\bibitem[Lin et~al\mbox{.}(2024a)]%
        {lin2024rtgs}
\bibfield{author}{\bibinfo{person}{Weikai Lin}, \bibinfo{person}{Yu Feng}, {and} \bibinfo{person}{Yuhao Zhu}.} \bibinfo{year}{2024}\natexlab{a}.
\newblock \showarticletitle{{RTGS: Enabling Real-Time Gaussian Splatting on Mobile Devices Using Efficiency-Guided Pruning and Foveated Rendering}}.
\newblock \bibinfo{journal}{\emph{arXiv preprint 2407.00435}} (\bibinfo{year}{2024}).
\newblock
\urldef\tempurl%
\url{https://arxiv.org/abs/2407.00435}
\showURL{%
\tempurl}


\bibitem[Mallick et~al\mbox{.}(2024)]%
        {mallick2024taming}
\bibfield{author}{\bibinfo{person}{Saswat Mallick}, \bibinfo{person}{Rahul Goel}, \bibinfo{person}{Bernhard Kerbl}, \bibinfo{person}{Francisco Vicente~Carrasco}, \bibinfo{person}{Markus Steinberger}, {and} \bibinfo{person}{Fernando De~La~Torre}.} \bibinfo{year}{2024}\natexlab{}.
\newblock \showarticletitle{T{aming 3DGS: High-Quality Radiance Fields with Limited Resources}}. In \bibinfo{booktitle}{\emph{SIGGRAPH Asia Conference Papers}}.
\newblock


\bibitem[Mantiuk et~al\mbox{.}(2012)]%
        {mantiuk2012comparison}
\bibfield{author}{\bibinfo{person}{Rafa{\l}~K Mantiuk}, \bibinfo{person}{Anna Tomaszewska}, {and} \bibinfo{person}{Rados{\l}aw Mantiuk}.} \bibinfo{year}{2012}\natexlab{}.
\newblock \showarticletitle{Comparison of four subjective methods for image quality assessment}.
\newblock \bibinfo{journal}{\emph{Computer Graphics Forum}} \bibinfo{volume}{31}, \bibinfo{number}{8} (\bibinfo{year}{2012}), \bibinfo{pages}{2478--2491}.
\newblock


\bibitem[Mildenhall et~al\mbox{.}(2020)]%
        {Mildenhall2020NeRF}
\bibfield{author}{\bibinfo{person}{Ben Mildenhall}, \bibinfo{person}{Pratul~P. Srinivasan}, \bibinfo{person}{Matthew Tancik}, \bibinfo{person}{Jonathan~T. Barron}, \bibinfo{person}{Ravi Ramamoorthi}, {and} \bibinfo{person}{Ren Ng}.} \bibinfo{year}{2020}\natexlab{}.
\newblock \showarticletitle{{NeRF: Representing Scenes as Neural Radiance Fields for View Synthesis}}. In \bibinfo{booktitle}{\emph{European Conference on Computer Vision}}.
\newblock


\bibitem[Moenne-Loccoz et~al\mbox{.}(2024)]%
        {moenneloccoz2024gaussianraytracing}
\bibfield{author}{\bibinfo{person}{Nicolas Moenne-Loccoz}, \bibinfo{person}{Ashkan Mirzaei}, \bibinfo{person}{Riccardo Perel, or de~Lutio}, \bibinfo{person}{Janick Martinez~Esturo}, \bibinfo{person}{Gavriel State}, \bibinfo{person}{Sanja Fidler}, \bibinfo{person}{Nicholas Sharo}, {and} \bibinfo{person}{Zan Gojcic}.} \bibinfo{year}{2024}\natexlab{}.
\newblock \showarticletitle{{3D Gaussian Ray Tracing: Fast Tracing of Particle Scenes}}.
\newblock \bibinfo{journal}{\emph{ACM TOG}} \bibinfo{volume}{43}, \bibinfo{number}{6} (\bibinfo{year}{2024}).
\newblock


\bibitem[Mueller and Crawfis(1998)]%
        {mueller1998popping}
\bibfield{author}{\bibinfo{person}{Klaus Mueller} {and} \bibinfo{person}{Roger Crawfis}.} \bibinfo{year}{1998}\natexlab{}.
\newblock \showarticletitle{{Eliminating Popping Artifacts in Sheet Buffer-Based Splatting}}. In \bibinfo{booktitle}{\emph{IEEE Visualization}}.
\newblock


\bibitem[M\"uller et~al\mbox{.}(2022)]%
        {mueller2022instant}
\bibfield{author}{\bibinfo{person}{Thomas M\"uller}, \bibinfo{person}{Alex Evans}, \bibinfo{person}{Christoph Schied}, {and} \bibinfo{person}{Alexander Keller}.} \bibinfo{year}{2022}\natexlab{}.
\newblock \showarticletitle{{Instant Neural Graphics Primitives with a Multiresolution Hash Encoding}}.
\newblock \bibinfo{journal}{\emph{ACM TOG}} \bibinfo{volume}{41}, \bibinfo{number}{4}, Article \bibinfo{articleno}{102} (\bibinfo{year}{2022}), \bibinfo{numpages}{15}~pages.
\newblock


\bibitem[Nguyen-Phuoc et~al\mbox{.}(2022)]%
        {nguyen2022snerf}
\bibfield{author}{\bibinfo{person}{Thu Nguyen-Phuoc}, \bibinfo{person}{Feng Liu}, {and} \bibinfo{person}{Lei Xiao}.} \bibinfo{year}{2022}\natexlab{}.
\newblock \showarticletitle{{SNeRF: Stylized Neural Implicit Representations for 3D Scenes}}.
\newblock \bibinfo{journal}{\emph{ACM TOG}} \bibinfo{volume}{41}, \bibinfo{number}{4} (\bibinfo{year}{2022}), \bibinfo{pages}{142:1--142:11}.
\newblock


\bibitem[Niemeyer et~al\mbox{.}(2024)]%
        {niemeyer2024radsplat}
\bibfield{author}{\bibinfo{person}{Michael Niemeyer}, \bibinfo{person}{Fabian Manhardt}, \bibinfo{person}{Marie-Julie Rakotosaona}, \bibinfo{person}{Michael Oechsle}, \bibinfo{person}{Daniel Duckworth}, \bibinfo{person}{Rama Gosula}, \bibinfo{person}{Keisuke Tateno}, \bibinfo{person}{John Bates}, \bibinfo{person}{Dominik Kaeser}, {and} \bibinfo{person}{Federico Tombari}.} \bibinfo{year}{2024}\natexlab{}.
\newblock \showarticletitle{{RadSplat: Radiance Field-Informed Gaussian Splatting for Robust Real-Time Rendering with 900+ FPS}}.
\newblock \bibinfo{journal}{\emph{arXiv preprint 2403.13806}} (\bibinfo{year}{2024}).
\newblock
\urldef\tempurl%
\url{https://arxiv.org/abs/2403.13806}
\showURL{%
\tempurl}


\bibitem[Patney et~al\mbox{.}(2016)]%
        {patney2016foveated}
\bibfield{author}{\bibinfo{person}{Anjul Patney}, \bibinfo{person}{Marco Salvi}, \bibinfo{person}{Joohwan Kim}, \bibinfo{person}{Anton Kaplanyan}, \bibinfo{person}{Chris Wyman}, \bibinfo{person}{Nir Benty}, \bibinfo{person}{David Luebke}, {and} \bibinfo{person}{Aaron Lefohn}.} \bibinfo{year}{2016}\natexlab{}.
\newblock \showarticletitle{{Towards Foveated Rendering for Gaze-Tracked Virtual Reality}}.
\newblock \bibinfo{journal}{\emph{ACM TOG}} \bibinfo{volume}{35}, \bibinfo{number}{6}, Article \bibinfo{articleno}{179} (\bibinfo{year}{2016}), \bibinfo{numpages}{12}~pages.
\newblock


\bibitem[Radl et~al\mbox{.}(2024a)]%
        {radl2024laenerf}
\bibfield{author}{\bibinfo{person}{Lukas Radl}, \bibinfo{person}{Michael Steiner}, \bibinfo{person}{Andreas Kurz}, {and} \bibinfo{person}{Markus Steinberger}.} \bibinfo{year}{2024}\natexlab{a}.
\newblock \showarticletitle{{LAENeRF: Local Appearance Editing for Neural Radiance Fields}}. In \bibinfo{booktitle}{\emph{IEEE/CVF Conference on Computer Vision and Pattern Recognition}}.
\newblock


\bibitem[Radl et~al\mbox{.}(2024b)]%
        {radl2024stopthepop}
\bibfield{author}{\bibinfo{person}{Lukas Radl}, \bibinfo{person}{Michael Steiner}, \bibinfo{person}{Mathias Parger}, \bibinfo{person}{Alexander Weinrauch}, \bibinfo{person}{Bernhard Kerbl}, {and} \bibinfo{person}{Markus Steinberger}.} \bibinfo{year}{2024}\natexlab{b}.
\newblock \showarticletitle{{StopThePop: Sorted Gaussian Splatting for View-Consistent Real-time Rendering}}.
\newblock \bibinfo{journal}{\emph{ACM TOG}} \bibinfo{volume}{43}, \bibinfo{number}{4}, Article \bibinfo{articleno}{64} (\bibinfo{year}{2024}).
\newblock


\bibitem[Rolff et~al\mbox{.}(2023a)]%
        {rolff2023interactive}
\bibfield{author}{\bibinfo{person}{Tim Rolff}, \bibinfo{person}{Ke Li}, \bibinfo{person}{Julia Hertel}, \bibinfo{person}{Susanne Schmidt}, \bibinfo{person}{Simone Frintrop}, {and} \bibinfo{person}{Frank Steinicke}.} \bibinfo{year}{2023}\natexlab{a}.
\newblock \showarticletitle{Interactive VRS-NeRF: Lightning fast Neural Radiance Field Rendering for Virtual Reality}. In \bibinfo{booktitle}{\emph{Proceedings of the ACM Symposium on Spatial User Interaction}}.
\newblock


\bibitem[Rolff et~al\mbox{.}(2023b)]%
        {rolff2023vrs}
\bibfield{author}{\bibinfo{person}{Tim Rolff}, \bibinfo{person}{Susanne Schmidt}, \bibinfo{person}{Ke Li}, \bibinfo{person}{Frank Steinicke}, {and} \bibinfo{person}{Simone Frintrop}.} \bibinfo{year}{2023}\natexlab{b}.
\newblock \showarticletitle{VRS-NeRF: Accelerating Neural Radiance Field Rendering with Variable Rate Shading}. In \bibinfo{booktitle}{\emph{Proceedings of the IEEE International Symposium on Mixed and Augmented Reality}}. IEEE.
\newblock


\bibitem[Steiner et~al\mbox{.}(2024)]%
        {steiner2024nerfcaching}
\bibfield{author}{\bibinfo{person}{Michael Steiner}, \bibinfo{person}{Thomas Köhler}, \bibinfo{person}{Lukas Radl}, {and} \bibinfo{person}{Markus Steinberger}.} \bibinfo{year}{2024}\natexlab{}.
\newblock \showarticletitle{{Frustum Volume Caching for Accelerated NeRF Rendering}}.
\newblock \bibinfo{journal}{\emph{Proceedings of the ACM on Computer Graphics and Interactive Technologies}} \bibinfo{volume}{7}, \bibinfo{number}{3}, Article \bibinfo{articleno}{39} (\bibinfo{year}{2024}), \bibinfo{numpages}{22}~pages.
\newblock


\bibitem[Sun et~al\mbox{.}(2017)]%
        {sun2017perceptually}
\bibfield{author}{\bibinfo{person}{Qi Sun}, \bibinfo{person}{Fu-Chung Huang}, \bibinfo{person}{Joohwan Kim}, \bibinfo{person}{Li-Yi Wei}, \bibinfo{person}{David Luebke}, {and} \bibinfo{person}{Arie Kaufman}.} \bibinfo{year}{2017}\natexlab{}.
\newblock \showarticletitle{Perceptually-Guided Foveation for Light Field Displays}.
\newblock \bibinfo{journal}{\emph{ACM TOG}} \bibinfo{volume}{36}, \bibinfo{number}{6} (\bibinfo{year}{2017}), \bibinfo{pages}{1--13}.
\newblock


\bibitem[Tariq et~al\mbox{.}(2022)]%
        {tariy2022noisebased}
\bibfield{author}{\bibinfo{person}{Taimoor Tariq}, \bibinfo{person}{Cara Tursun}, {and} \bibinfo{person}{Piotr Didyk}.} \bibinfo{year}{2022}\natexlab{}.
\newblock \showarticletitle{{Noise-based Enhancement for Foveated Rendering}}.
\newblock \bibinfo{journal}{\emph{ACM TOG}} \bibinfo{volume}{41}, \bibinfo{number}{4}, Article \bibinfo{articleno}{143} (\bibinfo{year}{2022}), \bibinfo{numpages}{14}~pages.
\newblock


\bibitem[Tu et~al\mbox{.}(2024)]%
        {anon2024fastandrobust}
\bibfield{author}{\bibinfo{person}{Xuechang Tu}, \bibinfo{person}{Bernhard Kerbl}, {and} \bibinfo{person}{Fernando de~la Torre}.} \bibinfo{year}{2024}\natexlab{}.
\newblock \showarticletitle{{Fast and Robust 3D Gaussian Splatting for Virtual Reality}}. In \bibinfo{booktitle}{\emph{SIGGRAPH Asia Posters}}.
\newblock


\bibitem[Xu et~al\mbox{.}(2023)]%
        {xu_vr-nerf_2023}
\bibfield{author}{\bibinfo{person}{Linning Xu}, \bibinfo{person}{Vasu Agrawal}, \bibinfo{person}{William Laney}, \bibinfo{person}{Tony Garcia}, \bibinfo{person}{Aayush Bansal}, \bibinfo{person}{Changil Kim}, \bibinfo{person}{Samuel Rota~Bulò}, \bibinfo{person}{Lorenzo Porzi}, \bibinfo{person}{Peter Kontschieder}, \bibinfo{person}{Aljaž Božič}, \bibinfo{person}{Dahua Lin}, \bibinfo{person}{Michael Zollhöfer}, {and} \bibinfo{person}{Christian Richardt}.} \bibinfo{year}{2023}\natexlab{}.
\newblock \showarticletitle{{VR}-{NeRF}: {High}-{Fidelity} {Virtualized} {Walkable} {Spaces}}. In \bibinfo{booktitle}{\emph{SIGGRAPH Asia Conference Papers}}.
\newblock


\bibitem[Yu et~al\mbox{.}(2024)]%
        {Yu2023MipSplatting}
\bibfield{author}{\bibinfo{person}{Zehao Yu}, \bibinfo{person}{Anpei Chen}, \bibinfo{person}{Binbin Huang}, \bibinfo{person}{Torsten Sattler}, {and} \bibinfo{person}{Andreas Geiger}.} \bibinfo{year}{2024}\natexlab{}.
\newblock \showarticletitle{{Mip-Splatting: Alias-free 3D Gaussian Splatting}}.
\newblock \bibinfo{journal}{\emph{IEEE/CVF Conference on Computer Vision and Pattern Recognition}} (\bibinfo{year}{2024}).
\newblock


\bibitem[Zhang et~al\mbox{.}(2018)]%
        {zhang2018unreasonable}
\bibfield{author}{\bibinfo{person}{Richard Zhang}, \bibinfo{person}{Phillip Isola}, \bibinfo{person}{Alexei~A Efros}, \bibinfo{person}{Eli Shechtman}, {and} \bibinfo{person}{Oliver Wang}.} \bibinfo{year}{2018}\natexlab{}.
\newblock \showarticletitle{{The Unreasonable Effectiveness of Deep Features as a Perceptual Metric}}. In \bibinfo{booktitle}{\emph{IEEE/CVF Conference on Computer Vision and Pattern Recognition}}.
\newblock


\bibitem[Zhang et~al\mbox{.}(2023)]%
        {zhang2023refnpr}
\bibfield{author}{\bibinfo{person}{Yuechen Zhang}, \bibinfo{person}{Zexin He}, \bibinfo{person}{Jinbo Xing}, \bibinfo{person}{Xufeng Yao}, {and} \bibinfo{person}{Jiaya Jia}.} \bibinfo{year}{2023}\natexlab{}.
\newblock \showarticletitle{{Ref-NPR: Reference-Based Non-Photorealistic Radiance Fields for Controllable Scene Stylization}}. In \bibinfo{booktitle}{\emph{IEEE/CVF Conference on Computer Vision and Pattern Recognition}}.
\newblock


\bibitem[Zwicker et~al\mbox{.}(2001)]%
        {zwicker_ewa_2001}
\bibfield{author}{\bibinfo{person}{Mathias Zwicker}, \bibinfo{person}{Hanspeter Pfister}, \bibinfo{person}{Jeroen van Baar}, {and} \bibinfo{person}{Markus Gross}.} \bibinfo{year}{2001}\natexlab{}.
\newblock \showarticletitle{{EWA Volume Splatting}}. In \bibinfo{booktitle}{\emph{IEEE Visualization}}.
\newblock


\end{thebibliography}

\appendix

\section{Two-pass Foveated Rendering}
\label{app:twopassfovea}

\begin{figure}[!h]
	\centering
	\includegraphics[width=0.9\columnwidth, trim=0cm 9cm 1cm 3cm, clip]{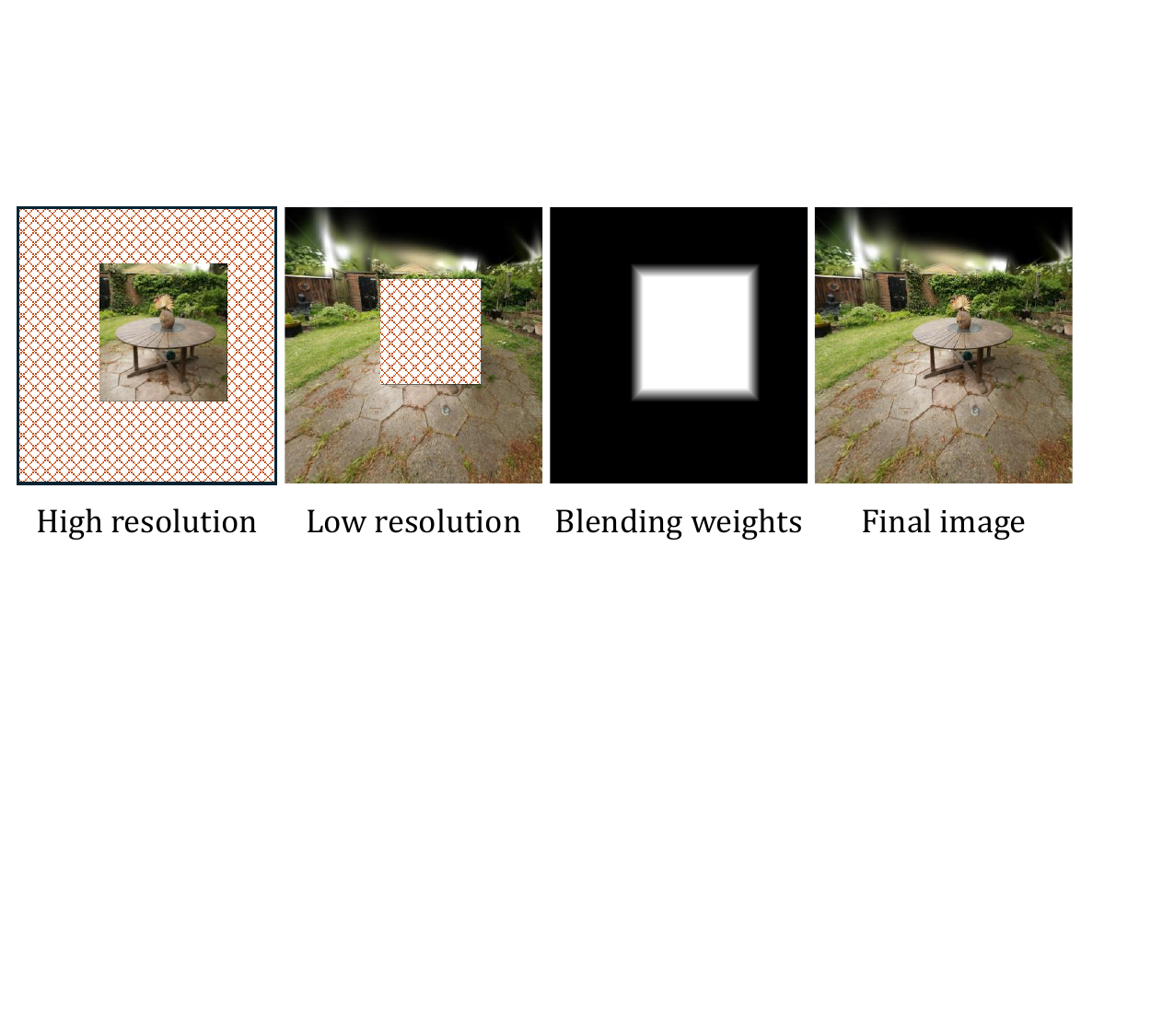}
	\caption{
Depiction of our simple two-pass foveated rendering solution:
we render the periphery at \revised{half}{a lower} resolution and the center \revised{at}{a} full resolution, performing blending at the boundary of the visibility mask~\cite{anon2024fastandrobust}.}
	\label{fig:two_pass_fovea}
\end{figure}

In addition to our proposed single-pass foveated renderer, we provide a simple, but optimized two-pass foveated rendering approach for comparison.
The first pass renders the center region at full resolution, using a tight projection frustum to maximize culling efficiency.
The second pass only renders peripheral content at a lower resolution, also considering the visibility mask as provided by OpenXR (see Sec.~\ref{sec:single_pass_foveated}).

To guarantee a smooth transition, we render content inside a narrow transition band twice, bilinearly upsample the low resolution content using NVIDIA Performance Primitives, and blend the two render pass images based on the continuous blending mask (see Fig. \ref{fig:two_pass_fovea}).
\new{The size of the center and transitional region for this setup is identical to Ours.}
This provides a strong baseline for our optimized single-pass foveated rendering scheme.

\section{User Study Details}
\label{app:userstudy}

In this section, we detail complimentary demographic information about our set of participants and provide detailed, per-scene user study results.

\paragraph{Demographic Survey.}
Our set of participants for the user study was comprised of 25 participants, aged 23-57.
All but one participant had normal or corrected vision:
this participant exhibited color blindness.
Note that 5 participants did not use their visual aid during the user study.

The majority of the tested population were familiar with computer graphics (3-5 on a 5 point Likert scale):
the remaining 4 participants ranked their knowledge of computer graphics as \say{very poor} or \say{poor}.
9 participants had never used a VR headset before, the remainder described their usage as \say{Yearly} up to \say{Daily}.
While 6 participants reported mild motion sickness during the user study, 20/25 ranked the overall experience as \say{Very satisfying} to \say{Completely satisfying}.
Each user study lasted between $20$ and $30$ minutes.

\begin{figure}[ht!]
    \centering
    \includegraphics[width=.8\linewidth]{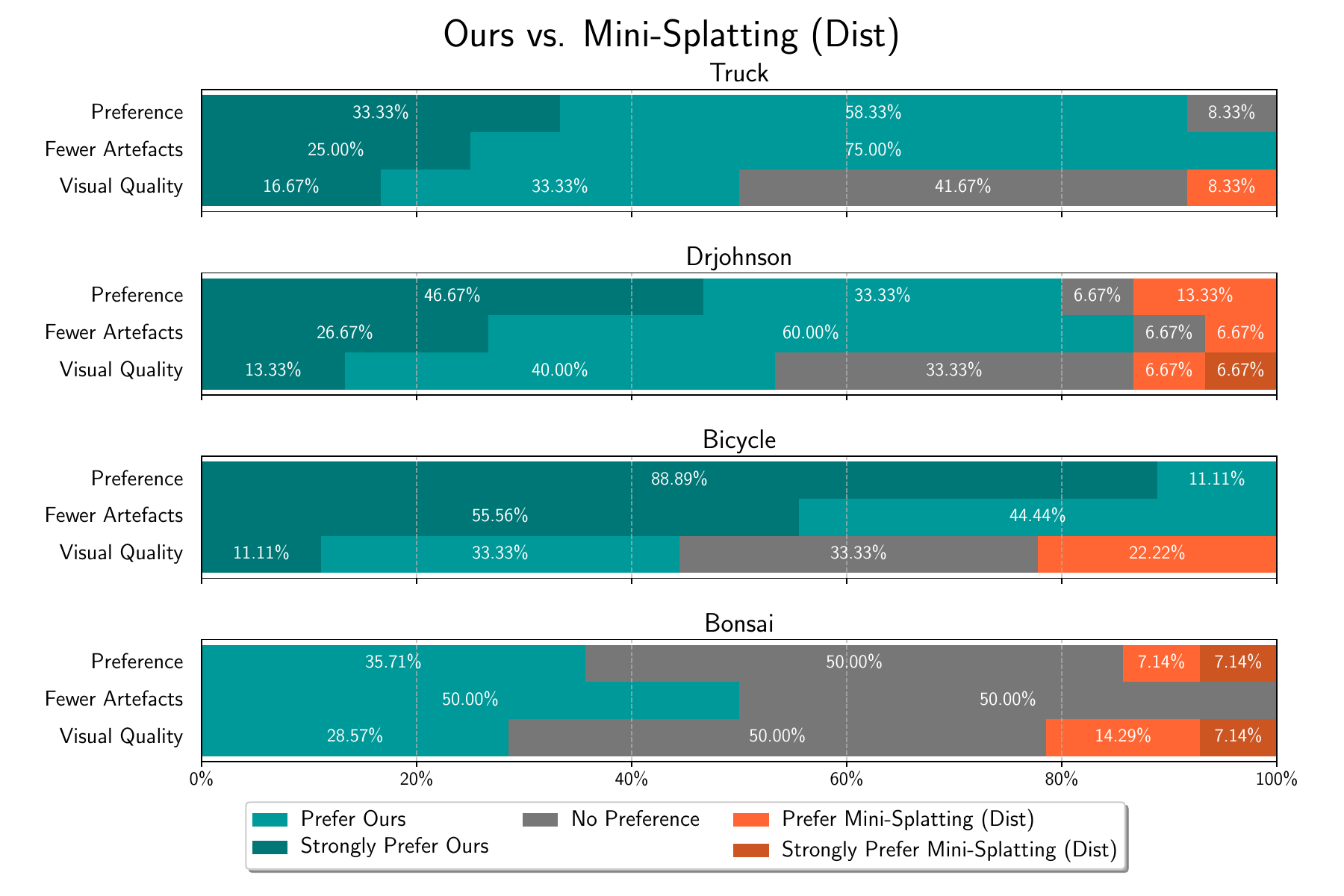} \\
    \includegraphics[width=.8\linewidth]{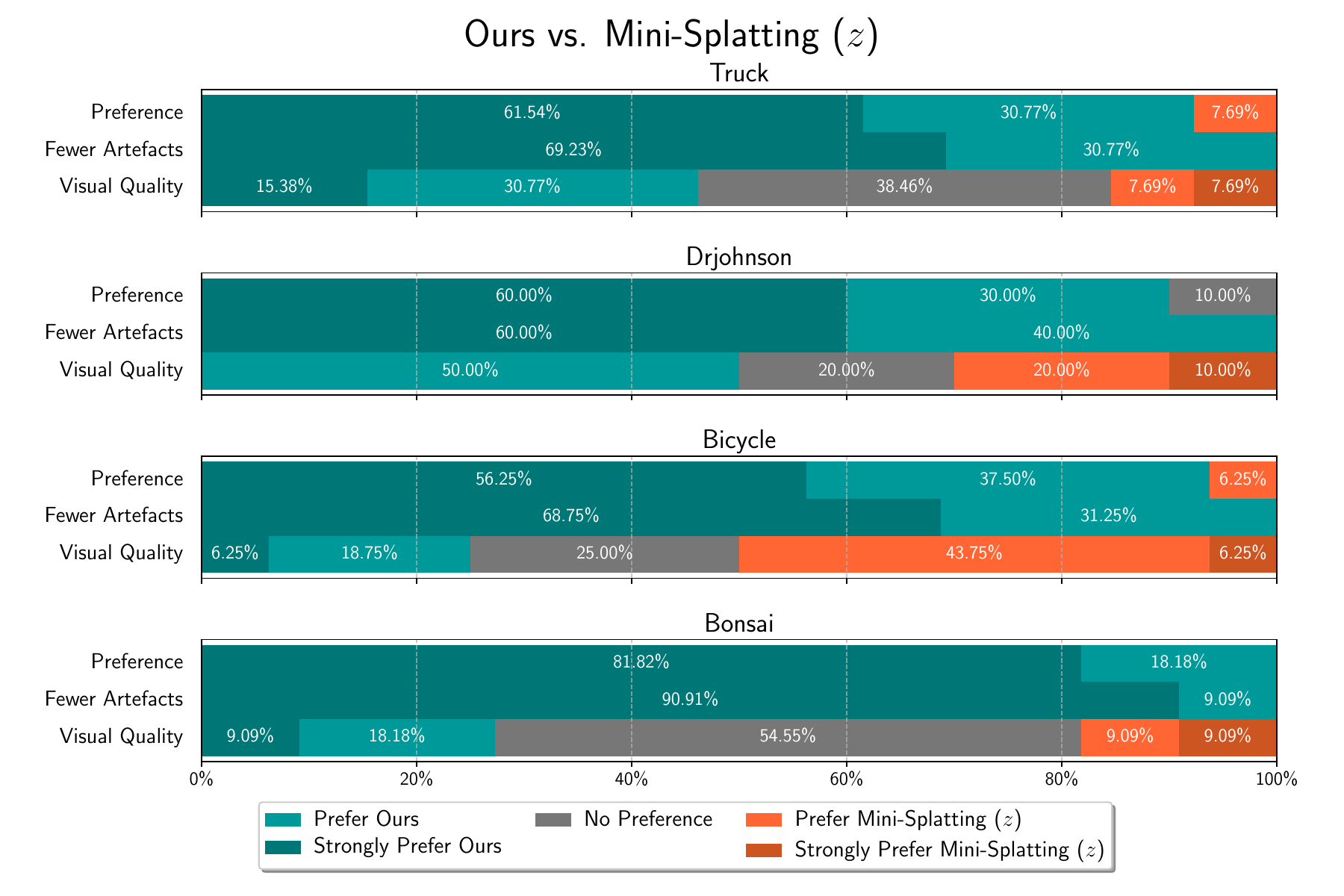}
    \caption{
Per-Scene user study results:
Depending on the scene, the prevalence of errors due to projection or popping artifacts can vary significantly.
Despite this, our method still performs favorably for each configuration.
}
    \label{fig:app:userstudy}
\end{figure}

\paragraph{Per-Scene results.}
The prevalence of various artifacts within 3DGS is scene-dependent:
therefore, we provide per-scene user study results in Fig.~\ref{fig:app:userstudy}.
As can clearly be seen, there is no clear outlier where Mini-Splatting outperforms our method to a large degree.
However, we find results for \emph{Bonsai} vs. Mini-Splatting (Dist) are the least significant out all of tested scenarios.
This stems from the fact that artifacts due to projection are less prevalent in this scene (see the supplemental video) and that popping artifacts are significantly reduced when sorting with Euclidean distance.

\section{Details on the Fine-Tuning Ablation}
\label{sec:app:finetuning}
In Tab.~\ref{tab:app:ablationstudy}, we provide detailed results for the ablation study presented in Fig.~\ref{fig:ablation_train}.
As can be seen, the scores for other image quality metrics generally align with the PSNR results.

For SSIM and LPIPS, the top-3 methods (highlighted) do not include Mini-Splatting:
however, StopThePop generally performs very well for these metrics, as can also be seen in Tab.~\ref{tab:app:metrics}.

\begin{table}[ht!]
    \centering
        \caption{Detailed results for our fine-tuning ablation study:
we provide image metrics, averaged over all tested scenes.
Clearly, StopThePop fine-tuning is able to recover image quality quickly despite the modified sort order.
}
    \small
    \begin{tabular}{lcccc}
    \toprule
        Method & PSNR\textsuperscript{$\uparrow$} & SSIM\textsuperscript{$\uparrow$} & LPIPS\textsuperscript{$\downarrow$} & \FLIP\textsuperscript{$\downarrow$} \\\midrule
        Ours (no fine-tuning) & \cellcolor{tab_color!0}25.458 & \cellcolor{tab_color!0}0.795 & \cellcolor{tab_color!0}0.254 & \cellcolor{tab_color!0}0.184 \\
        Ours $1\text{K}$ &  \cellcolor{tab_color!0}26.860 & \cellcolor{tab_color!0}0.832 & \cellcolor{tab_color!0}0.221 & \cellcolor{tab_color!0}0.153 \\
        Ours $2.5\text{K}$ & \cellcolor{tab_color!0}26.950 & \cellcolor{tab_color!10}0.835 & \cellcolor{tab_color!10}0.219 & \cellcolor{tab_color!0}0.151 \\
        Ours $5\text{K}$ & \cellcolor{tab_color!10}27.039 & \cellcolor{tab_color!30}0.836 & \cellcolor{tab_color!30}0.217 & \cellcolor{tab_color!10}0.150 \\
        Ours $10\text{K}$ & \cellcolor{tab_color!30}27.087 & \cellcolor{tab_color!50}0.837 & \cellcolor{tab_color!50}0.215 & \cellcolor{tab_color!30}0.149 \\\midrule
        Mini-Splatting & \cellcolor{tab_color!50}27.094 & \cellcolor{tab_color!0}0.835 & \cellcolor{tab_color!0}0.222 & \cellcolor{tab_color!50}0.146 \\
    \bottomrule
    \end{tabular}

    \label{tab:app:ablationstudy}
\end{table}

\section{Large FOV Evaluation Protocol}

\begin{figure}
    \centering
    \subcaptionbox{Ours (Dist)}[0.7\linewidth]{\includegraphics[width=\linewidth]{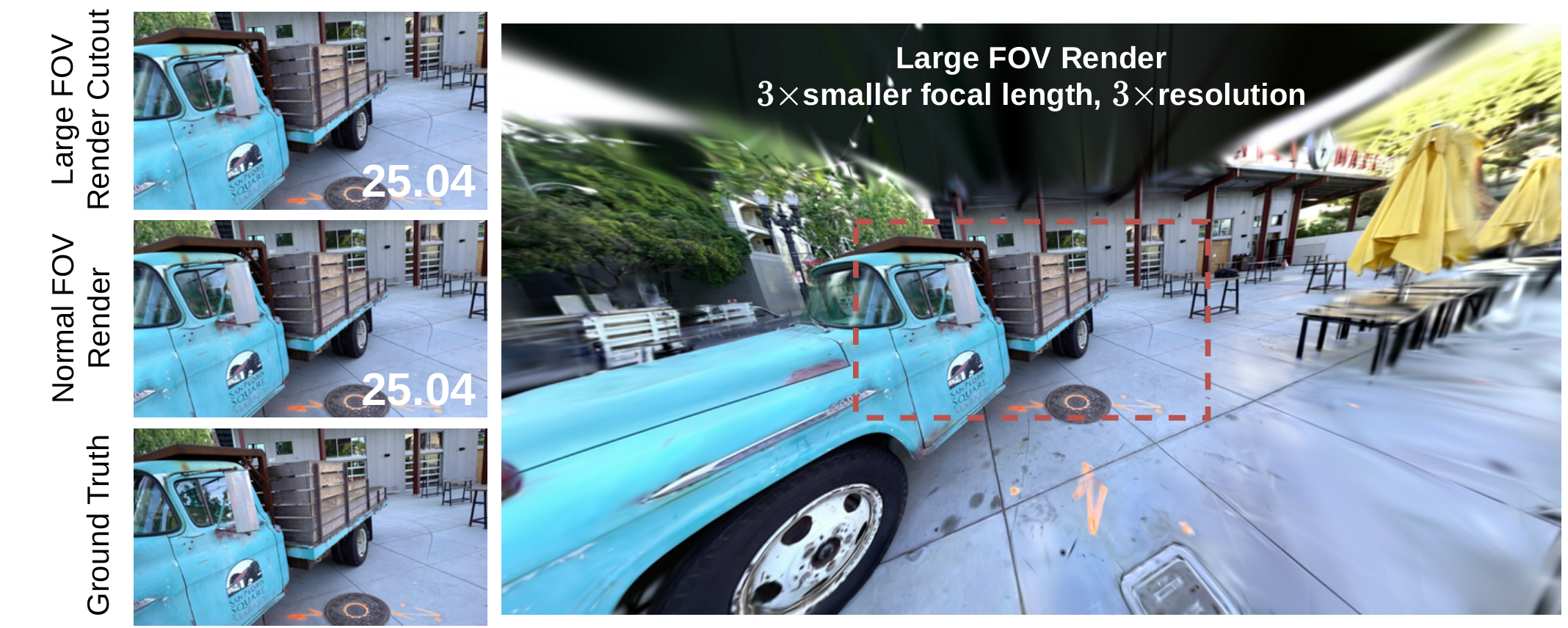}} \\
    \subcaptionbox{Mini-Splatting (Dist)}[0.7\linewidth]{\includegraphics[width=\linewidth]{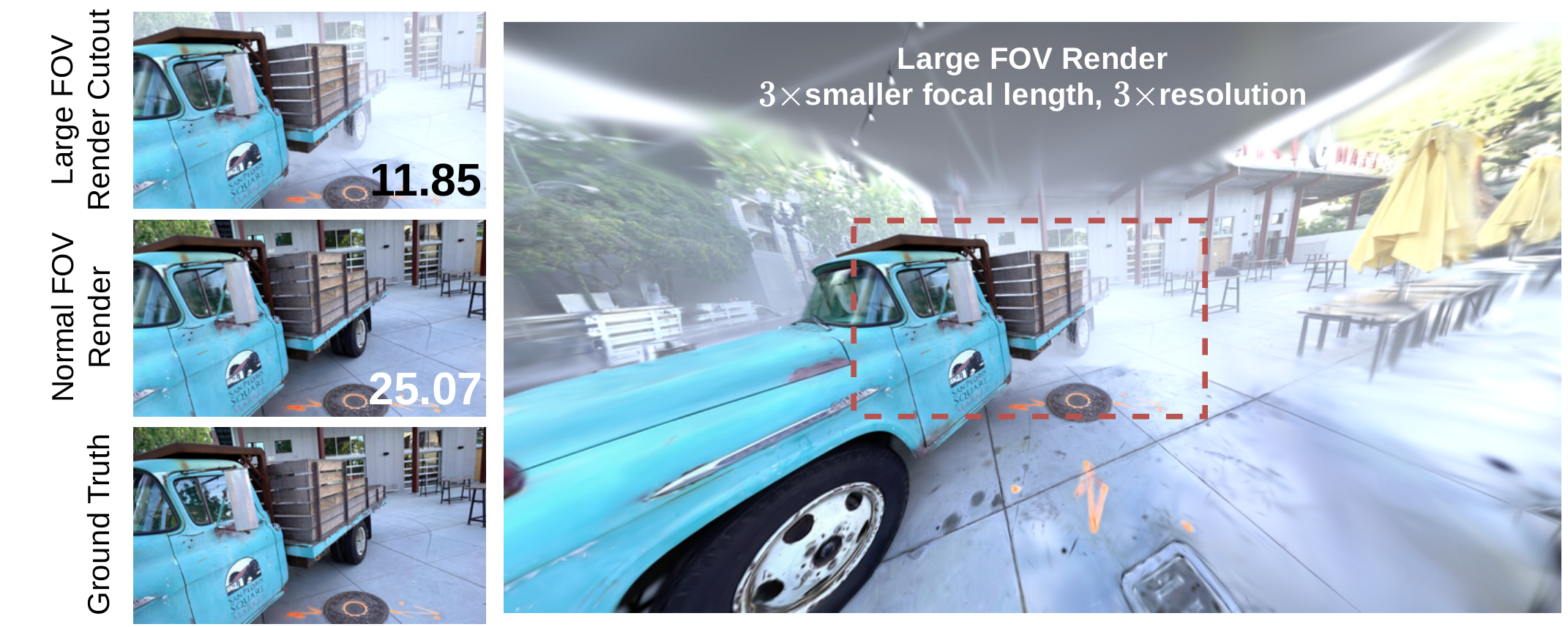}}
    \caption{Visualization of our large FOV evaluation setup, which more closely resembles the VR scenario and allows for comparison against ground truth test images (see inset PSNR values) via a pixel-perfect cutout.
Ours does not suffer from projection errors, while Mini-Splatting exhibits very strong artifacts.}
    \label{fig:app:large_fov_setup}
\end{figure}

\new{As previously discussed, we evaluate the robustness of our approach to projection artifacts in a VR setting, by employing a large FOV evaluation.
\citet{huang2024optimal} evaluate this by changing focal lengths to $0.2$ and $0.3$ times the original, cut out a patch that corresponds to the original focal length, and resizing the ground truth image for comparison.
We instead chose to decrease the original focal length by a factor of 3 and increase render resolution by 3 (see Fig.~\ref{fig:app:large_fov_setup}).
This has several advantages: (1) the center crop has the exact same resolution as the original render; (2) we receive a pixel-perfect cutout; (3) the dilation of 2D Gaussians (making them at least 1 pixel wide) does not change.
This also prevents any unwanted influence on image metrics due to interpolation during resizing or cropping; Hence, for a method without any projection errors the cutout should exactly match the image rendered with the original focal length.
In Fig.~\ref{fig:app:large_fov_setup}, the inset PSNR metrics show that this is the case for our method, but not for Mini-Splatting (as well as 3DGS and StopThePop), where projection errors lead to disturbing artifacts and much worse image metrics.
}

\section{Per-Scene Large FOV Image Metrics}

In Tab.~\ref{tab:app:perscene_largefov}, we provide accompanying per-scene results for the averaged results of Tab.~\ref{tab:image_metrics_large_fov}.

\begin{table*}[]
    \centering
        \caption{Detailed per-scene image metrics for evaluation with large FOV for all tested scenes.}\setlength{\tabcolsep}{4pt}
    \scriptsize
    \setlength{\tabcolsep}{2pt}
    \begin{tabular}{lccccccccccccc}
\toprule
Dataset & \multicolumn{5}{c}{Mip-NeRF 360 Outdoor} &\multicolumn{4}{c}{Mip-NeRF 360 Indoor} & \multicolumn{2}{c}{Deep Blending} & \multicolumn{2}{c}{Tanks \& Temples}\\
Scene & Bicycle & Flowers & Garden & Stump & Treehill & Bonsai & Counter & Kitchen & Room & DrJ & Playroom & Train & Truck\\\midrule
& \\[-2.4ex]
 & \multicolumn{13}{c}{PSNR\textsuperscript{$\uparrow$}} \\\cmidrule(lr){2-14}
3DGS & \cellcolor{tab_color!0} 25.08 & \cellcolor{tab_color!0} 20.09 & \cellcolor{tab_color!50} 27.10 & \cellcolor{tab_color!0} 26.34 & \cellcolor{tab_color!0} 20.98 & \cellcolor{tab_color!30} 31.78 & \cellcolor{tab_color!50} 28.92 & \cellcolor{tab_color!0} 30.73 & \cellcolor{tab_color!10} 30.37 & \cellcolor{tab_color!0} 26.92 & \cellcolor{tab_color!0} 25.46 & \cellcolor{tab_color!0} 15.92 & \cellcolor{tab_color!0} 18.30 \\
StopThePop & \cellcolor{tab_color!10} 25.21 & \cellcolor{tab_color!10} 21.29 & \cellcolor{tab_color!30} 27.01 & \cellcolor{tab_color!0} 26.70 & \cellcolor{tab_color!10} 22.22 & \cellcolor{tab_color!50} 31.80 & \cellcolor{tab_color!0} 28.57 & \cellcolor{tab_color!30} 31.03 & \cellcolor{tab_color!0} 29.53 & \cellcolor{tab_color!10} 27.52 & \cellcolor{tab_color!10} 27.58 & \cellcolor{tab_color!10} 19.53 & \cellcolor{tab_color!10} 20.95\\
Mini-Splatting ($z$) & \cellcolor{tab_color!0} 23.94 & \cellcolor{tab_color!0} 17.64 & \cellcolor{tab_color!0} 25.29 & \cellcolor{tab_color!10} 27.11 & \cellcolor{tab_color!0} 20.76 & \cellcolor{tab_color!0} 29.34 & \cellcolor{tab_color!0} 27.30 & \cellcolor{tab_color!0} 28.77 & \cellcolor{tab_color!0} 22.74 & \cellcolor{tab_color!0} 23.46 & \cellcolor{tab_color!0} 25.24 & \cellcolor{tab_color!0} 14.63 & \cellcolor{tab_color!0} 16.08\\
Mini-Splatting (Dist) & \cellcolor{tab_color!0} 24.16 & \cellcolor{tab_color!0} 19.10 & \cellcolor{tab_color!0} 25.49 & \cellcolor{tab_color!0} 27.06 & \cellcolor{tab_color!0} 20.80 & \cellcolor{tab_color!0} 29.85 & \cellcolor{tab_color!0} 28.32 & \cellcolor{tab_color!0} 29.89 & \cellcolor{tab_color!0} 23.69 & \cellcolor{tab_color!0} 25.31 & \cellcolor{tab_color!0} 26.95 & \cellcolor{tab_color!0} 16.74 & \cellcolor{tab_color!0} 17.29\\
Ours ($z$) & \cellcolor{tab_color!50} 25.28 & \cellcolor{tab_color!30} 21.57 & \cellcolor{tab_color!10} 26.77 & \cellcolor{tab_color!30} 27.19 & \cellcolor{tab_color!30} 22.76 & \cellcolor{tab_color!0} 31.40 & \cellcolor{tab_color!30} 28.65 & \cellcolor{tab_color!50} 31.18 & \cellcolor{tab_color!50} 30.85 & \cellcolor{tab_color!30} 29.81 & \cellcolor{tab_color!50} 30.85 & \cellcolor{tab_color!50} 21.20 & \cellcolor{tab_color!50} 24.61\\
Ours (Dist) & \cellcolor{tab_color!30} 25.27 & \cellcolor{tab_color!50} 21.67 & \cellcolor{tab_color!0} 26.76 & \cellcolor{tab_color!50} 27.23 & \cellcolor{tab_color!50} 22.79 & \cellcolor{tab_color!10} 31.46 & \cellcolor{tab_color!10} 28.62 & \cellcolor{tab_color!10} 30.74 & \cellcolor{tab_color!30} 30.83 & \cellcolor{tab_color!50} 29.85 & \cellcolor{tab_color!30} 30.82 & \cellcolor{tab_color!30} 21.18 & \cellcolor{tab_color!30} 24.50\\
\midrule
& \\[-2.4ex]
 & \multicolumn{13}{c}{SSIM\textsuperscript{$\uparrow$}} \\\cmidrule(lr){2-14}
3DGS & \cellcolor{tab_color!0} 0.759 & \cellcolor{tab_color!0} 0.571 & \cellcolor{tab_color!50} 0.864 & \cellcolor{tab_color!0} 0.768 & \cellcolor{tab_color!0} 0.598 & \cellcolor{tab_color!30} 0.940 & \cellcolor{tab_color!50} 0.907 & \cellcolor{tab_color!0} 0.923 & \cellcolor{tab_color!0} 0.902 & \cellcolor{tab_color!0} 0.883 & \cellcolor{tab_color!0} 0.866 & \cellcolor{tab_color!0} 0.692 & \cellcolor{tab_color!0} 0.790\\
StopThePop & \cellcolor{tab_color!10} 0.766 & \cellcolor{tab_color!10} 0.599 & \cellcolor{tab_color!30} 0.863 & \cellcolor{tab_color!0} 0.775 & \cellcolor{tab_color!10} 0.631 & \cellcolor{tab_color!50} 0.940 & \cellcolor{tab_color!0} 0.904 & \cellcolor{tab_color!10} 0.925 & \cellcolor{tab_color!10} 0.903 & \cellcolor{tab_color!10} 0.889 & \cellcolor{tab_color!10} 0.890 & \cellcolor{tab_color!10} 0.776 & \cellcolor{tab_color!10} 0.842\\
Mini-Splatting ($z$) & \cellcolor{tab_color!0} 0.748 & \cellcolor{tab_color!0} 0.526 & \cellcolor{tab_color!0} 0.838 & \cellcolor{tab_color!0} 0.804 & \cellcolor{tab_color!0} 0.615 & \cellcolor{tab_color!0} 0.931 & \cellcolor{tab_color!0} 0.897 & \cellcolor{tab_color!0} 0.915 & \cellcolor{tab_color!0} 0.798 & \cellcolor{tab_color!0} 0.864 & \cellcolor{tab_color!0} 0.878 & \cellcolor{tab_color!0} 0.651 & \cellcolor{tab_color!0} 0.722\\
Mini-Splatting (Dist) & \cellcolor{tab_color!0} 0.742 & \cellcolor{tab_color!0} 0.564 & \cellcolor{tab_color!0} 0.835 & \cellcolor{tab_color!10} 0.804 & \cellcolor{tab_color!0} 0.623 & \cellcolor{tab_color!0} 0.931 & \cellcolor{tab_color!0} 0.903 & \cellcolor{tab_color!0} 0.921 & \cellcolor{tab_color!0} 0.824 & \cellcolor{tab_color!0} 0.872 & \cellcolor{tab_color!0} 0.888 & \cellcolor{tab_color!0} 0.723 & \cellcolor{tab_color!0} 0.785 \\
Ours ($z$) & \cellcolor{tab_color!50} 0.775 & \cellcolor{tab_color!50} 0.628 & \cellcolor{tab_color!10} 0.845 & \cellcolor{tab_color!50} 0.808 & \cellcolor{tab_color!50} 0.654 & \cellcolor{tab_color!10} 0.939 & \cellcolor{tab_color!30} 0.906 & \cellcolor{tab_color!50} 0.927 & \cellcolor{tab_color!50} 0.921 & \cellcolor{tab_color!30} 0.908 & \cellcolor{tab_color!50} 0.915 & \cellcolor{tab_color!30} 0.798 & \cellcolor{tab_color!50} 0.874\\
Ours (Dist) & \cellcolor{tab_color!30} 0.774 & \cellcolor{tab_color!30} 0.628 & \cellcolor{tab_color!0} 0.843 & \cellcolor{tab_color!30} 0.807 & \cellcolor{tab_color!30} 0.653 & \cellcolor{tab_color!0} 0.939 & \cellcolor{tab_color!10} 0.906 & \cellcolor{tab_color!30} 0.926 & \cellcolor{tab_color!30} 0.921 & \cellcolor{tab_color!50} 0.909 & \cellcolor{tab_color!30} 0.915 & \cellcolor{tab_color!50} 0.798 & \cellcolor{tab_color!30} 0.872 \\
\midrule
& \\[-2.4ex]
 & \multicolumn{13}{c}{LPIPS\textsuperscript{$\downarrow$}} \\\cmidrule(lr){2-14}
3DGS & \cellcolor{tab_color!30} 0.213 & \cellcolor{tab_color!0} 0.353 & \cellcolor{tab_color!30} 0.107 & \cellcolor{tab_color!0} 0.217 & \cellcolor{tab_color!0} 0.338 & \cellcolor{tab_color!0} 0.200 & \cellcolor{tab_color!0} 0.198 & \cellcolor{tab_color!10} 0.126 & \cellcolor{tab_color!0} 0.217 & \cellcolor{tab_color!0} 0.247 & \cellcolor{tab_color!0} 0.246 & \cellcolor{tab_color!0} 0.264 & \cellcolor{tab_color!0} 0.194\\
StopThePop & \cellcolor{tab_color!50} 0.205 & \cellcolor{tab_color!10} 0.335 & \cellcolor{tab_color!50} 0.106 & \cellcolor{tab_color!0} 0.209 & \cellcolor{tab_color!10} 0.319 & \cellcolor{tab_color!0} 0.200 & \cellcolor{tab_color!0} 0.197 & \cellcolor{tab_color!50} 0.125 & \cellcolor{tab_color!10} 0.216 & \cellcolor{tab_color!50} 0.243 & \cellcolor{tab_color!0} 0.244 & \cellcolor{tab_color!50} 0.218 & \cellcolor{tab_color!10} 0.166\\
Mini-Splatting ($z$) & \cellcolor{tab_color!0} 0.231 & \cellcolor{tab_color!0} 0.396 & \cellcolor{tab_color!0} 0.150 & \cellcolor{tab_color!0} 0.198 & \cellcolor{tab_color!0} 0.325 & \cellcolor{tab_color!0} 0.198 & \cellcolor{tab_color!0} 0.198 & \cellcolor{tab_color!0} 0.130 & \cellcolor{tab_color!0} 0.237 & \cellcolor{tab_color!0} 0.274 & \cellcolor{tab_color!0} 0.249 & \cellcolor{tab_color!0} 0.309 & \cellcolor{tab_color!0} 0.258\\
Mini-Splatting (Dist) & \cellcolor{tab_color!0} 0.235 & \cellcolor{tab_color!0} 0.364 & \cellcolor{tab_color!0} 0.154 & \cellcolor{tab_color!10} 0.196 & \cellcolor{tab_color!0} 0.327 & \cellcolor{tab_color!10} 0.197 & \cellcolor{tab_color!10} 0.196 & \cellcolor{tab_color!0} 0.130 & \cellcolor{tab_color!0} 0.227 & \cellcolor{tab_color!0} 0.268 & \cellcolor{tab_color!10} 0.242 & \cellcolor{tab_color!0} 0.277 & \cellcolor{tab_color!0} 0.202\\
Ours ($z$) & \cellcolor{tab_color!10} 0.218 & \cellcolor{tab_color!30} 0.322 & \cellcolor{tab_color!10} 0.148 & \cellcolor{tab_color!30} 0.195 & \cellcolor{tab_color!50} 0.308 & \cellcolor{tab_color!30} 0.196 & \cellcolor{tab_color!50} 0.193 & \cellcolor{tab_color!30} 0.126 & \cellcolor{tab_color!30} 0.206 & \cellcolor{tab_color!10} 0.246 & \cellcolor{tab_color!50} 0.239 & \cellcolor{tab_color!30} 0.238 & \cellcolor{tab_color!50} 0.157\\
Ours (Dist) & \cellcolor{tab_color!0} 0.220 & \cellcolor{tab_color!50} 0.322 & \cellcolor{tab_color!0} 0.150 & \cellcolor{tab_color!50} 0.194 & \cellcolor{tab_color!30} 0.310 & \cellcolor{tab_color!50} 0.195 & \cellcolor{tab_color!30} 0.193 & \cellcolor{tab_color!0} 0.127 & \cellcolor{tab_color!50} 0.205 & \cellcolor{tab_color!30} 0.245 & \cellcolor{tab_color!30} 0.239 & \cellcolor{tab_color!10} 0.239 & \cellcolor{tab_color!30} 0.158\\
\bottomrule \\
\end{tabular}

    \label{tab:app:perscene_largefov}
\end{table*}

\section{Standard Image Metrics}
\label{app:image_metrics}

\new{We provide per-scene results for our standard image metrics evaluation (original focal length) in Tab.~\ref{tab:app:perscene_new}, as well as results averaged over all datasets in Tab.~\ref{tab:app:metrics}.}

\begin{table*}[]
    \centering
        \caption{Standard image quality metrics (without changes in focal length): our proposed method performs well on a variety of datasets. Additionally, our image quality metrics rival those of Mini-Splatting~\cite{fang2024minisplatting}, even though only 5K fine-tuning iterations were used}\setlength{\tabcolsep}{4pt}
    \footnotesize
\begin{tabular}{lrrrrrrrrr}
\toprule
Dataset & \multicolumn{3}{c}{Mip-NeRF 360} & \multicolumn{3}{c}{Tanks \& Temples} & \multicolumn{3}{c}{Deep Blending}\\
\cmidrule(lr){2-4} \cmidrule(lr){5-7} \cmidrule(lr){8-10}
 & PSNR\textsuperscript{$\uparrow$} & SSIM\textsuperscript{$\uparrow$} & LPIPS\textsuperscript{$\downarrow$} & PSNR\textsuperscript{$\uparrow$} & SSIM\textsuperscript{$\uparrow$} & LPIPS\textsuperscript{$\downarrow$} & PSNR\textsuperscript{$\uparrow$} & SSIM\textsuperscript{$\uparrow$} & LPIPS\textsuperscript{$\downarrow$}  \\
\midrule
StopThePop & \cellcolor{tab_color!10} 27.304 & 0.815 & \cellcolor{tab_color!50} 0.211 & \cellcolor{tab_color!10} 23.226 & \cellcolor{tab_color!30} 0.846 & \cellcolor{tab_color!50} 0.171 & 29.929 & 0.908 & \cellcolor{tab_color!50} 0.231 \\
3DGS & \cellcolor{tab_color!50} 27.443 & 0.814 & 0.215 & \cellcolor{tab_color!50} 23.734 & \cellcolor{tab_color!50} 0.847 & \cellcolor{tab_color!30} 0.175 & 29.510 & 0.902 & \cellcolor{tab_color!30} 0.237 \\
\midrule
Mini-Splatting ($z$) & \cellcolor{tab_color!30} 27.326 & 0.821 & 0.215 & \cellcolor{tab_color!30} 23.308 & 0.835 & 0.200 & 29.994 & 0.907 & 0.243 \\
Mini-Splatting (Dist) & 27.227 & \cellcolor{tab_color!10} 0.821 & 0.214 & 23.084 & 0.834 & 0.198 & \cellcolor{tab_color!10} 30.148 & \cellcolor{tab_color!10} 0.909 & \cellcolor{tab_color!10} 0.239 \\
Ours ($z$) & 27.294 & \cellcolor{tab_color!50} 0.823 & \cellcolor{tab_color!30} 0.212 & 22.901 & \cellcolor{tab_color!10} 0.836 & \cellcolor{tab_color!10} 0.197 & \cellcolor{tab_color!30} 30.333 & \cellcolor{tab_color!30} 0.912 & 0.242 \\
Ours (Dist) & 27.263 & \cellcolor{tab_color!30} 0.822 & \cellcolor{tab_color!10} 0.213 & 22.840 & 0.835 & 0.198 & \cellcolor{tab_color!50} 30.337 & \cellcolor{tab_color!50} 0.912 & 0.242 \\
\bottomrule
\end{tabular}

    \label{tab:app:metrics}
\end{table*}

\begin{table*}[]
    \centering
        \caption{\new{Detailed per-scene image metrics for all tested scenes (original focal length).}}\setlength{\tabcolsep}{4pt}
    \scriptsize
    \setlength{\tabcolsep}{1.5pt}
    \begin{tabular}{lccccccccccccc}
\toprule
Dataset & \multicolumn{5}{c}{Mip-NeRF 360 Outdoor} &\multicolumn{4}{c}{Mip-NeRF 360 Indoor} & \multicolumn{2}{c}{Deep Blending} & \multicolumn{2}{c}{Tanks \& Temples} \\
Scene & Bicycle & Flowers & Garden & Stump & Treehill & Bonsai & Counter & Kitchen & Room & DrJ & Playroom & Train & Truck\\\midrule
& \\[-2.4ex]
 & \multicolumn{13}{c}{PSNR\textsuperscript{$\uparrow$}} \\\cmidrule(lr){2-14}
3DGS & \cellcolor{tab_color!0} 25.19 & \cellcolor{tab_color!0} 21.53 & \cellcolor{tab_color!50} 27.30 & \cellcolor{tab_color!0} 26.62 & \cellcolor{tab_color!0} 22.46 & \cellcolor{tab_color!50} 32.11 & \cellcolor{tab_color!50} 28.97 & \cellcolor{tab_color!50} 31.33 & \cellcolor{tab_color!50} 31.48 & \cellcolor{tab_color!0} 29.05 & \cellcolor{tab_color!0} 29.97 & \cellcolor{tab_color!50} 22.05 & \cellcolor{tab_color!50} 25.41\\
StopThePop & \cellcolor{tab_color!0} 25.22 & \cellcolor{tab_color!0} 21.54 & \cellcolor{tab_color!30} 27.23 & \cellcolor{tab_color!0} 26.70 & \cellcolor{tab_color!0} 22.44 & \cellcolor{tab_color!30} 31.98 & \cellcolor{tab_color!0} 28.60 & \cellcolor{tab_color!10} 31.18 & \cellcolor{tab_color!0} 30.84 & \cellcolor{tab_color!0} 29.45 & \cellcolor{tab_color!0} 30.40 & \cellcolor{tab_color!10} 21.49 & \cellcolor{tab_color!10} 24.96\\
Mini-Splatting ($z$) & \cellcolor{tab_color!0} 25.21 & \cellcolor{tab_color!0} 21.46 & \cellcolor{tab_color!0} 26.89 & \cellcolor{tab_color!10} 27.21 & \cellcolor{tab_color!0} 22.65 & \cellcolor{tab_color!10} 31.48 & \cellcolor{tab_color!0} 28.55 & \cellcolor{tab_color!0} 31.11 & \cellcolor{tab_color!30} 31.38 & \cellcolor{tab_color!0} 29.47 & \cellcolor{tab_color!0} 30.52 & \cellcolor{tab_color!30} 21.56 & \cellcolor{tab_color!30} 25.06\\
Mini-Splatting (Dist) & \cellcolor{tab_color!10} 25.22 & \cellcolor{tab_color!30} 21.61 & \cellcolor{tab_color!10} 26.91 & \cellcolor{tab_color!30} 27.22 & \cellcolor{tab_color!10} 22.69 & \cellcolor{tab_color!0} 31.35 & \cellcolor{tab_color!0} 28.52 & \cellcolor{tab_color!0} 30.68 & \cellcolor{tab_color!10} 30.84 & \cellcolor{tab_color!10} 29.64 & \cellcolor{tab_color!10} 30.66 & \cellcolor{tab_color!0} 21.33 & \cellcolor{tab_color!0} 24.84\\
Ours ($z$) & \cellcolor{tab_color!50} 25.28 & \cellcolor{tab_color!10} 21.57 & \cellcolor{tab_color!0} 26.77 & \cellcolor{tab_color!0} 27.19 & \cellcolor{tab_color!30} 22.76 & \cellcolor{tab_color!0} 31.40 & \cellcolor{tab_color!30} 28.65 & \cellcolor{tab_color!30} 31.18 & \cellcolor{tab_color!0} 30.84 & \cellcolor{tab_color!30} 29.81 & \cellcolor{tab_color!50} 30.85 & \cellcolor{tab_color!0} 21.20 & \cellcolor{tab_color!0} 24.61 \\
Ours (Dist) & \cellcolor{tab_color!30} 25.27 & \cellcolor{tab_color!50} 21.67 & \cellcolor{tab_color!0} 26.76 & \cellcolor{tab_color!50} 27.23 & \cellcolor{tab_color!50} 22.79 & \cellcolor{tab_color!0} 31.45 & \cellcolor{tab_color!10} 28.62 & \cellcolor{tab_color!0} 30.74 & \cellcolor{tab_color!0} 30.83 & \cellcolor{tab_color!50} 29.85 & \cellcolor{tab_color!30} 30.82 & \cellcolor{tab_color!0} 21.18 & \cellcolor{tab_color!0} 24.50\\
\midrule
& \\[-2.4ex]
 & \multicolumn{13}{c}{SSIM\textsuperscript{$\uparrow$}} \\\cmidrule(lr){2-14}
3DGS & \cellcolor{tab_color!0} 0.764 & \cellcolor{tab_color!0} 0.605 & \cellcolor{tab_color!50} 0.864 & \cellcolor{tab_color!0} 0.772 & \cellcolor{tab_color!0} 0.633 & \cellcolor{tab_color!30} 0.941 & \cellcolor{tab_color!50} 0.907 & \cellcolor{tab_color!0} 0.926 & \cellcolor{tab_color!0} 0.919 & \cellcolor{tab_color!0} 0.900 & \cellcolor{tab_color!0} 0.905 & \cellcolor{tab_color!50} 0.814 & \cellcolor{tab_color!30} 0.880\\
StopThePop & \cellcolor{tab_color!0} 0.768 & \cellcolor{tab_color!0} 0.605 & \cellcolor{tab_color!30} 0.864 & \cellcolor{tab_color!0} 0.776 & \cellcolor{tab_color!0} 0.635 & \cellcolor{tab_color!50} 0.941 & \cellcolor{tab_color!0} 0.905 & \cellcolor{tab_color!30} 0.926 & \cellcolor{tab_color!0} 0.918 & \cellcolor{tab_color!10} 0.906 & \cellcolor{tab_color!0} 0.910 & \cellcolor{tab_color!30} 0.810 & \cellcolor{tab_color!50} 0.882\\
Mini-Splatting ($z$) & \cellcolor{tab_color!0} 0.772 & \cellcolor{tab_color!0} 0.625 & \cellcolor{tab_color!0} 0.847 & \cellcolor{tab_color!0} 0.805 & \cellcolor{tab_color!0} 0.653 & \cellcolor{tab_color!0} 0.939 & \cellcolor{tab_color!0} 0.904 & \cellcolor{tab_color!0} 0.924 & \cellcolor{tab_color!10} 0.921 & \cellcolor{tab_color!0} 0.903 & \cellcolor{tab_color!0} 0.910 & \cellcolor{tab_color!0} 0.798 & \cellcolor{tab_color!0} 0.873\\
Mini-Splatting (Dist) & \cellcolor{tab_color!10} 0.773 & \cellcolor{tab_color!10} 0.626 & \cellcolor{tab_color!10} 0.847 & \cellcolor{tab_color!10} 0.805 & \cellcolor{tab_color!10} 0.653 & \cellcolor{tab_color!0} 0.939 & \cellcolor{tab_color!0} 0.904 & \cellcolor{tab_color!0} 0.923 & \cellcolor{tab_color!0} 0.920 & \cellcolor{tab_color!0} 0.905 & \cellcolor{tab_color!10} 0.912 & \cellcolor{tab_color!0} 0.796 & \cellcolor{tab_color!0} 0.873\\
Ours ($z$) & \cellcolor{tab_color!50} 0.775 & \cellcolor{tab_color!50} 0.628 & \cellcolor{tab_color!0} 0.845 & \cellcolor{tab_color!50} 0.808 & \cellcolor{tab_color!50} 0.654 & \cellcolor{tab_color!10} 0.939 & \cellcolor{tab_color!30} 0.906 & \cellcolor{tab_color!50} 0.927 & \cellcolor{tab_color!50} 0.921 & \cellcolor{tab_color!30} 0.908 & \cellcolor{tab_color!50} 0.915 & \cellcolor{tab_color!0} 0.798 & \cellcolor{tab_color!10} 0.874\\
Ours (Dist) & \cellcolor{tab_color!30} 0.774 & \cellcolor{tab_color!30} 0.628 & \cellcolor{tab_color!0} 0.843 & \cellcolor{tab_color!30} 0.807 & \cellcolor{tab_color!30} 0.653 & \cellcolor{tab_color!0} 0.939 & \cellcolor{tab_color!10} 0.906 & \cellcolor{tab_color!10} 0.926 & \cellcolor{tab_color!30} 0.921 & \cellcolor{tab_color!50} 0.909 & \cellcolor{tab_color!30} 0.915 & \cellcolor{tab_color!10} 0.798 & \cellcolor{tab_color!0} 0.872\\
\midrule
& \\[-2.4ex]
 & \multicolumn{13}{c}{LPIPS\textsuperscript{$\downarrow$}} \\\cmidrule(lr){2-14}
3DGS & \cellcolor{tab_color!30} 0.210 & \cellcolor{tab_color!0} 0.335 & \cellcolor{tab_color!30} 0.107 & \cellcolor{tab_color!0} 0.214 & \cellcolor{tab_color!0} 0.326 & \cellcolor{tab_color!0} 0.200 & \cellcolor{tab_color!0} 0.198 & \cellcolor{tab_color!30} 0.125 & \cellcolor{tab_color!0} 0.216 & \cellcolor{tab_color!30} 0.240 & \cellcolor{tab_color!10} 0.234 & \cellcolor{tab_color!30} 0.205 & \cellcolor{tab_color!30} 0.144\\
StopThePop & \cellcolor{tab_color!50} 0.204 & \cellcolor{tab_color!0} 0.332 & \cellcolor{tab_color!50} 0.106 & \cellcolor{tab_color!0} 0.208 & \cellcolor{tab_color!0} 0.316 & \cellcolor{tab_color!0} 0.199 & \cellcolor{tab_color!0} 0.197 & \cellcolor{tab_color!50} 0.125 & \cellcolor{tab_color!0} 0.214 & \cellcolor{tab_color!50} 0.231 & \cellcolor{tab_color!30} 0.231 & \cellcolor{tab_color!50} 0.202 & \cellcolor{tab_color!50} 0.140\\
Mini-Splatting ($z$) & \cellcolor{tab_color!0} 0.222 & \cellcolor{tab_color!0} 0.326 & \cellcolor{tab_color!10} 0.147 & \cellcolor{tab_color!0} 0.197 & \cellcolor{tab_color!10} 0.313 & \cellcolor{tab_color!30} 0.195 & \cellcolor{tab_color!0} 0.195 & \cellcolor{tab_color!0} 0.128 & \cellcolor{tab_color!0} 0.208 & \cellcolor{tab_color!0} 0.252 & \cellcolor{tab_color!0} 0.235 & \cellcolor{tab_color!0} 0.243 & \cellcolor{tab_color!0} 0.157\\
Mini-Splatting (Dist) & \cellcolor{tab_color!0} 0.222 & \cellcolor{tab_color!10} 0.324 & \cellcolor{tab_color!0} 0.147 & \cellcolor{tab_color!10} 0.195 & \cellcolor{tab_color!0} 0.313 & \cellcolor{tab_color!50} 0.194 & \cellcolor{tab_color!10} 0.195 & \cellcolor{tab_color!0} 0.129 & \cellcolor{tab_color!30} 0.206 & \cellcolor{tab_color!0} 0.248 & \cellcolor{tab_color!50} 0.230 & \cellcolor{tab_color!0} 0.241 & \cellcolor{tab_color!10} 0.155\\
Ours ($z$) & \cellcolor{tab_color!10} 0.218 & \cellcolor{tab_color!30} 0.322 & \cellcolor{tab_color!0} 0.148 & \cellcolor{tab_color!30} 0.195 & \cellcolor{tab_color!50} 0.308 & \cellcolor{tab_color!0} 0.196 & \cellcolor{tab_color!50} 0.193 & \cellcolor{tab_color!10} 0.126 & \cellcolor{tab_color!10} 0.206 & \cellcolor{tab_color!0} 0.246 & \cellcolor{tab_color!0} 0.239 & \cellcolor{tab_color!10} 0.238 & \cellcolor{tab_color!0} 0.157\\
Ours (Dist) & \cellcolor{tab_color!0} 0.220 & \cellcolor{tab_color!50} 0.322 & \cellcolor{tab_color!0} 0.150 & \cellcolor{tab_color!50} 0.194 & \cellcolor{tab_color!30} 0.310 & \cellcolor{tab_color!10} 0.195 & \cellcolor{tab_color!30} 0.193 & \cellcolor{tab_color!0} 0.127 & \cellcolor{tab_color!50} 0.205 & \cellcolor{tab_color!10} 0.245 & \cellcolor{tab_color!0} 0.239 & \cellcolor{tab_color!0} 0.239 & \cellcolor{tab_color!0} 0.158\\
\bottomrule \\
\end{tabular}

    \label{tab:app:perscene_new}
\end{table*}

\end{document}